\documentclass[fleqn,usenatbib]{mnras}

\usepackage{newtxtext,newtxmath}
 
\usepackage[T1]{fontenc}
\usepackage{ae,aecompl}

\usepackage{graphicx}	% Including figure files
\usepackage{amsmath}	% Advanced maths commands
\usepackage{amssymb}	% Extra maths symbols
\usepackage{dsfont}
\graphicspath{{Figures/}}

\title[GMC collisions' impact on star formation]{Towards the impact of GMC collisions on the star formation rate}

\author[G. H. Hunter et al.]{
Glen H. Hunter$^{1,2,3}$\thanks{E-mail: glen.hunter@uni-heidelberg.de},
Paul C. Clark$^{3}$,
Simon C. O. Glover$^{1}$,
Ralf S. Klessen$^{1,4}$
\\
$^{1}$Universit\"at Heidelberg, Zentrum f\"ur Astronomie, Institut f\"ur Theoretische Astrophysik, Albert-Ueberle-Str. 2, 69120 Heidelberg, Germany\\
$^{2}$SUPA, School of Physics \& Astronomy, University of St Andrews, North Haugh, St Andrews, KY16 9SS, United Kingdom\\
$^{3}$School of Physics and Astronomy, Queen's Buildings, The Parade, Cardiff University, Cardiff, CF24 3AA, United Kingdom\\
$^{4}$Universit\"at Heidelberg, Interdiszipli\"ares Zentrum f\"ur Wissenschaftliches Rechnen, Im Neuenheimer Feld 205, 69120 Heidelberg, Germany
}

\date{Accepted XXX. Received YYY; in original form ZZZ}

\pubyear{2022}

\begin{document}
\label{firstpage}
\pagerange{\pageref{firstpage}--\pageref{lastpage}}
\maketitle

Abstract of the paper
\begin{abstract}
Collisions between giant molecular clouds (GMCs) are one of the pathways for massive star formation, due to the high densities created. However the enhancement of the star formation rate (SFR) is not well constrained. In this study we perform a parameter study of cloud-cloud collisions, and investigate how the resulting SFR depends on the details of set-up.  Our parameter study explores variations in: collision speed; magnetic field inclination (with respect to the collisional axis); and resolution, as defined by the number of cells per Jeans length. In all our collision simulations we find a factor of 2-3 increase in the SFR compared to our no collision simulation, with star formation beginning sooner with a) high collisional velocities, b) parallel orientation between the magnetic field and collision axis, c) and lower resolution. The mean virial parameter of high density (and thus possible star-forming) gas increases with collisional velocity, but has little variation with magnetic field inclination. The alignment of the velocity and magnetic field remains uniform in low density environments but becomes more perpendicular with increasing density, indicating the compression of the magnetic field by collapsing gas.
Comparing the trends in the SFR with other GMC collision studies, we find good agreement with studies that account for the gravitational boundedness of the gas in their star formation algorithm, but not with those that simply form stars above a prescribed density threshold. This suggests that the latter approach should be used with caution when modelling star formation on resolved cloud scales.
\end{abstract}




















\begin{keywords}
stars: formation -- ISM: clouds -- galaxies: ISM
\end{keywords}

%%%%%%%%%%%%%%%%%%%%%%%%%%%%%%%%%%%%%%%%%%%%%%%%%%

%%%%%%%%%%%%%%%%% BODY OF PAPER %%%%%%%%%%%%%%%%%%
\section{Introduction}

Giant molecular clouds (GMCs) are large regions of gravitationally bound gas that exist within galaxies. These clouds typically span 1--100~pc in size and contain on the order $\sim 10^4$--$10^6$M$_\odot$ of molecular gas, giving them typical hydrogen densities of $\sim$100 cm$^{-3}$ \citep{Blitz1993,Duval2014}. GMCs have temperatures of $\sim10-30$~K, a factor of 2-10 times lower than the temperature of the surrounding Cold Neutral Medium (CNM) component of the interstellar medium (ISM). Due to their low temperatures and molecular nature, the main observational tracers of GMC properties are dust continuum emission and molecular line emission (frome e.g.\ CO/HCN/N$_2$H$^+$), which are used to observe GMCs within the Milky Way as well as in nearby galaxies \citep{1999PASJ...51..745F,2001ApJ...547..792D,2010A&A...518L.102A,Leroy2021}.

Embedded within these GMCs are cooler dense clumps, the more massive of which are often observed as Infrared Dark Cloud (IRDCs). These clumps are accepted to be the precursor to star formation \citep{2014prpl.conf..149T} due to their temperatures of 10-20~K \citep{2013A&A...552A..40C} and higher number densities of $n_{\mathrm{H}_2} = 10^4-10^6$cm$^{-3}$ \citep{2018ARA&A..56...41M}. Each clump will likely form a cluster of stars due to further fragmentation and their relatively large mass reservoir ($10^2$--$10^5$M$_\odot$) \citep{2018ARA&A..56...41M}. Despite many observations of these clumps, the process by which these clumps form is not fully understood. Some theoretical models have been proposed that explain how a GMC can fragment and further collapse; such as turbulence \citep{2005ApJ...630..250K}, stellar and supernova feedback \citep{2017MNRAS.470.2283W} and converging gas flows, i.e collisions \citep{1986ApJ...310L..77S,2017ApJ...841...88W}.

Numerical simulations of galactic disks have shown that GMC collisions are a recurring event due to the quasi-2D geometry of the distribution of dense molecular gas in most disk galaxies along with the differential rotation created by the gravitational potential of the galaxy \citep{2009ApJ...700..358T}. Estimates of the mean time between collisions range from $\sim 20$\% \citep{2015MNRAS.446.3608D} to $\sim 50$\% (Sun et al., in prep.) of the orbital period of the clouds around the centre of the galaxy. Although this is a factor of a few longer than recent observational estimates of cloud lifetimes -- between 10-30 Myr \citep{Chevance2020} -- it nevertheless implies that GMC collisions should be relatively common events. 
This is supported by the many observations of cloud collisions found within the Milky Way; for instance, \citet{2021PASJ...73S...1F} identify $\sim$50 GMCs that are candidates to be the product of cloud-cloud collisions (CCCs). However, collisions are not limited to the disks of galaxies. Cloud collisions have been observed in the Small and Large Magellanic Clouds (SMC and LMC) \citep{2019ApJ...886...14F,2019ApJ...886...15T,2021ApJ...908L..43N} and theorised to occur with extreme collisional velocities in the centre of the Milky Way \citep{2019MNRAS.488.4663S}. In the case of the SMC and LMC, these clouds collide as a result of large scale HI flows caused by the tidal interactions of these satellite galaxies \citep{2017PASJ...69L...5F}. As for the centre of the galaxy, \citet{2019MNRAS.488.4663S} argues the gas flow caused by the galactic bar crashes into the Central Molecular Zone, a ring of molecular gas surrounding the Milky Way nucleus, providing the ideal conditions for clouds to collide.

A variety of GMC collision simulations have been performed to explore various topics: such as observational signature of collisons \citep[e.g.][]{Haworth2015}; triggering of high-mass star formation \citep[e.g.][]{Takahira2014,Balfour2015}; and enhancement of the star formation rate, in both non-identical GMC collisions \citep[e.g.][]{1992PASJ...44..203H} and identical GMC collsions \citep[e.g.][]{2017ApJ...841...88W,Liow2020,2020MNRAS.494..246T}. \citet{1992PASJ...44..203H} and \citet{Takahira2014} show that the collision of a smaller cloud with a larger cloud creates a cavity and compression layer in which densities are reached that are high enough for high-mass stars to form. This is consistent with observations, for example of the Gum 11 cloud in the Carina Nebula Complex by \citet{2021PASJ...73S.201F}, where CO observations indicate the presence of colliding clouds with a cavity, and with a massive star, HD92206, in close proximity to the compressed layer.  

However, while there is general agreement that cloud-cloud collisions should enhance star formation, there is little agreement on {\em how much} the star formation rate is increased by a cloud-cloud collision.
Recent simulations by \citet{2020MNRAS.494..246T} and \citet{Liow2020} find that the star formation rate increases by a factor of two or less for collisions with relative velocities $v_{\rm rel} \leq 20 \: {\rm km \, s^{-1}}$, typical of the majority of mergers that we expect to occur in a Milky-Way type spiral galaxy \citep{Skarbinski2022}. On the other hand, \citet{2017ApJ...841...88W} find an order of magnitude increase in the star formation rate for a collision with a relative velocity of $10 \: {\rm km \, s^{-1}}$. The cause of the substantial discrepancy between the \citet{2017ApJ...841...88W} results and the results of the other cloud collision simulations is unclear, as the simulations differ in both their numerical approach\footnote{\citet{Liow2020} and \citet{2020MNRAS.494..246T} use smoothed particle hydrodynamics (SPH) and model star formation with sink particles, while \citet{2017ApJ...841...88W} use an Eulerian grid code and model star formation with star particles.
Both SPH studies adopt an isothermal equation of state, whereas \citet{2017ApJ...841...88W} employ a more realistic treatment of the heating and cooling that can occur during the collision, based on the microphysics of the interstellar medium.} and their initial conditions. In particular, \citet{2017ApJ...841...88W} include a magnetic field in many of their simulations, while the other studies consider only the purely hydrodynamical case.\footnote{Other potentially important differences include the choice of initial cloud mass and relative velocity. \citet{2020MNRAS.494..246T} consider an initial relative velocity of $10 \: {\rm km \, s^{-1}}$, as in W17, but simulate much smaller clouds, with masses of only $10^{4} \: {\rm M_{\odot}}$, compared to $\sim 10^{5} \: {\rm M_{\odot}}$ in W17. \citet{Liow2020} carry out simulations for a wide range of cloud masses, including one that is within a few percent of the W17 value, but only consider relative velocities $v_{\rm rel} \geq 20 \: {\rm km \, s^{-1}}$, significantly higher than the case studied in W17.} It is therefore not clear whether the difference in outcomes of these studies is a consequence of the different initial conditions adopted, or instead is a consequence of the choice of numerical approach.

In this paper, we attempt to improve our understanding of the effect of collisions on the star-formation rate in magnetised clouds by performing a series of simulations of cloud collisions with different relative velocities and magnetic field orientations using a state-of-the-art MHD code -- the {\sc arepo} moving-mesh code \citep{2010MNRAS.401..791S} -- and a sophisticated treatment of the microphysics of the gas (see \citealt{2019MNRAS.486.4622C} and our Appendix \ref{app:chem}) and the formation of stars (see \citealt{Wollenberg2020} and Section \ref{sect:SF}). Importantly, for our fiducial case, we adopt the same initial conditions as in \citeauthor{2017ApJ...841...88W}~(2017; hereafter, W17), allowing us to directly assess whether the large boost in the star formation rate that they find in their study is due to their choice of initial conditions or to their numerical approach. 

The structure of our paper as follows. In Section~\ref{sect:method} we discuss the numerical approach taken, the initial conditions used for each simulation and how the star formation is implemented. In Section~\ref{sect:results} we present our results from the simulation, with a focus on the structure of the GMC, the star formation rate and the virial parameter of the cloud. We conclude with a comparison of our results to other literature results in Section~\ref{sect:disc} and a summary of our findings in Section~\ref{sect:concl}. 

\section{Methodology}
\label{sect:method}

\subsection{Numerical magnetohydrodynamics}
\label{sect:MHD}

We make use of the \textsc{arepo} moving mesh code to simulate the gas dynamics and star formation of colliding GMCs  \citep{2010MNRAS.401..791S}. The code solves the equations of magnetohydrodynamics (MHD) \citep{10.1093/mnras/stt428},
\begin{equation}
    \label{eq:MHD1}
    \frac{\partial\rho}{\partial t} + \nabla \cdot (\rho\mathbf{v}) =  0 \,,
\end{equation}
\begin{equation}
    \label{eq:MHD2}
    \frac{\partial\rho\mathbf{v}}{\partial t} + \nabla \cdot (\rho\mathbf{v}\otimes\mathbf{v} + p_\text{tot}\mathds{1} + \mathbf{B}\otimes\mathbf{B}) = -\rho\nabla\Phi \,,
\end{equation}
\begin{equation}
    \label{eq:MHD3}
    \frac{\partial\rho e}{\partial t} + \nabla\cdot[\mathbf{v}(\rho e+p_\text{tot})-\mathbf{B}(\mathbf{v}\cdot\mathbf{B})] = \Dot{Q} + \rho\frac{\partial\Phi}{\partial t} \,,
\end{equation}
\begin{equation}
    \label{eq:MHD4}
    \frac{\partial\mathbf{B}}{\partial t} + \nabla \cdot (\mathbf{B}\otimes\mathbf{v} - \mathbf{v}\otimes\mathbf{B}) = 0 \,,
\end{equation}
where $\rho$, $\mathbf{v}$ and $\mathbf{B}$ are the density, velocity and magnetic field strength of a given cell. Here, $\mathds{1}$ is the identity matrix. The total pressure is the sum of the thermal and magnetic pressures, $p_\text{tot} = p_\text{gas} + \frac{1}{2}|\mathbf{B}|^2$. The total energy per unit mass is $e = e_\text{th} + \frac{1}{2}\mathbf{v}^2 + \Phi + \frac{1}{2\rho}|\mathbf{B}|^2$, where $e_{\rm th}$ is the thermal energy per unit mass. An adiabatic equation of state is adopted where $p_\text{gas} = (\gamma-1)\rho e_\text{th}$ with $\gamma = 5/3$. Heating and cooling of the gas due to chemical changes and radiative processes is accounted for with the term $\Dot{Q}$, which is discussed in more detail in Section~\ref{sect:chemistry}.

The equations are solved on a tessellated Voronoi mesh in which the mesh generating points are able to follow the gas flow using a Harten-Lax-van Leer discontinuity (HLLD) solver. This allows \textsc{arepo} to act as a quasi-Lagrangian MHD code. The Voronoi mesh is adaptive: cells can be refined or de-refined by adding or removing mesh-generating points, respectively. The divergence-free constraint on the magnetic field, $\nabla \cdot \mathbf{B} = 0$, is enforced by the using the MHD solver provided in \textsc{arepo} \citep{10.1093/mnras/stt428}. Here, additional source terms are added to Equations \ref{eq:MHD2}-\ref{eq:MHD4} following the scheme introduced by \cite{1999JCoPh.154..284P} combined with a hyperbolic Dedner cleaning \citep{DEDNER2002645}. This minimizes any effect that a diverging magnetic field may create. 

The gravitational term is due to the self-gravitation of the gas and any sink particles present within the system (see below). \textsc{arepo} solves Poisson's equation,
\begin{equation}\label{eq:poisson}
    \nabla^2\Phi = 4\pi\text{G}\rho \,,
\end{equation}
via a tree-based algorithm similar to the one used in the \textsc{gadget-2} code \citep{2005MNRAS.364.1105S}, where G is the gravitational constant. The algorithm treats each cell as if the mass is at a point in the centre of the cell with a degree of gravity softening. The softening length for the gas is adaptive and is given as $\varepsilon_\text{gas}=2r_\text{cell}$, where the $r_\text{cell}$ is the radius of a sphere with the same volume as the Voronoi cell. The minimum softening length for both the gas and sink particles is $40.02$~AU. Further information on sink particles will be discussed in Section~\ref{sect:SF}.

\subsection{Chemical network}
\label{sect:chemistry}
For our simulations we make use of a modified version of the chemical network developed by \citet{2017ApJ...843...38G}, which itself is an improved version of prior networks developed by \citet{1999ApJ...524..923N} and \cite{2012MNRAS.421..116G}. It includes a simplified treatment of the chemistry of H, C, and O and allows us to follow the evolution of the abundances of the main low temperature gas coolants (CO, C, O, and C$^{+}$) with high accuracy, but low computational cost. A comprehensive description of the network can be found in \citet{2017ApJ...843...38G}, and full details of the modifications we have made to it are described in Appendix~\ref{app:chem}.
Radiative heating and cooling of the gas are modelled using a detailed atomic and molecular cooling function, most recently described in \cite{2019MNRAS.486.4622C}.

To treat the effects of H$_{2}$, C and CO self-shielding as well as shielding by dust, we make use of the {\sc Treecol} algorithm \citep{2012MNRAS.420..745C}. It uses information stored in the gravitational tree structure to compute a $4\pi$ steradian map of the column densities of each of these species plus dust around each Voronoi cell. These maps are then used to determine how the interstellar radiation field (ISRF) reaching the cell is attenuated by self-shielding and dust absorption. 

In all of our simulations, we adopt elemental carbon and oxygen abundances given by \citet{2000ApJ...528..310S}. Following \citet{1978ApJS...36..595D}, we set the strength of the interstellar radiation field (ISRF) to $G_0 = 1.7$ in \citet{1968BAN....19..421H} units. Finally, we adopt a value of $\zeta_{\rm H} = 3 \times10^{-17} \: {\rm s^{-1}}$ for the cosmic ray ionisation rate of atomic hydrogen.

\subsection{Star formation}
\label{sect:SF}
To simulate star formation, we make use of sink particles (henceforth sinks) to represent forming star clusters \cite[][]{1995MNRAS.277..362B,2010ApJ...713..269F}. We follow the sink creation protocol outlined in \citet{Wollenberg2020}. To summarise, the following conditions must be satisfied in order for a sink to be created:
\begin{enumerate}
\item[a)] The cell must have a density greater than a threshold density $\rho_\text{c} = 1.991\times10^{-16}$ g cm$^{-3}$. This threshold has been motivated by the work of \citet{Jones_etal_2022}, who demonstrate that only collapsing -- and thus actively star-forming -- gas can reach such densities in such a set-up. \citet{Prole_etal_2022} show that provided that the sink particles form within the collapsing regime, the star formation rate is insensitive to the exact choice of the threshold density.  Note that while we could insert sink particles at lower densities, this reduces the efficiency of the algorithm, as it needs to check more candidates for sink creation. Also, this would increase the chance of converting gas to sinks, that while {\em currently} bound, may be subsequently disrupted via further interactions with the large-scale flows.

\item[b)] The gas flow within the accretion radius of the sink ($r_\text{acc} = 187$~AU, corresponding to the Jeans length at the threshold density for a temperature of 10~K) must be converging. We ensure this by requiring convergence of both the velocity field ($\nabla\cdot\mathbf{v} < 0$) and the acceleration field ($\nabla\cdot\mathbf{a}<0$).

\item[c)] The sink-forming region must be situated within a local minimum of the gravitational potential.

\item[d)] The cell must not lie within a distance $r < r_{\rm acc}$ of an existing sink particle.

\item[e)] The region within $r_{\rm acc}$ must be gravitationally bound, i.e $|E_\text{grav}| > 2(E_\text{therm}+E_\text{kin})$ where $E_\text{grav}$ is the gravitational energy, and $E_\text{therm}$ and $E_\text{kin}$ are the thermal and kinetic energies, respectfully.
\end{enumerate}

If all of these criteria are met, the gas cell is converted into a sink that inherits its mass and momentum. This sink is able to interact gravitationally with the surrounding environment and is also able to accrete further gas onto itself. Any Voronoi cells with $\rho > \rho_\text{c}$ within a distance $r < r_\text{acc}$ of a sink are potentially eligible for accretion. However, gas is only accreted from the cell if it is gravitationally bound to the sink. Provided this is the case, enough gas is removed from the cell to reduce its density to $\rho_{\rm c}$, although the total amount removed at each time-step is capped at 90\% of the cell's initial mass, for reasons of numerical stability. Following the accretion, any properties of the cell that depend on its mass are updated. It should be noted that the sinks formed do not contribute to the magnetic field of the system (i.e $|B_\text{sink}| = 0$). Given the size of our sink particles, and the density at which they are introduced, it is clear that they cannot be interpreted as individual stars, but rather as small-$N$ protostellar systems. Also, given that there is no feedback from the young stars in our cloud, any star formation efficiencies reported in this paper should be considered as upper limits -- our sink particles record the mass that is capable of going into stars (i.e.,\ trapped in potential wells), in accordance with the above star formation model.

\subsection{Initial conditions}
\label{sect:ICs}
\begin{figure}
    \centering
    \includegraphics[width=\columnwidth]{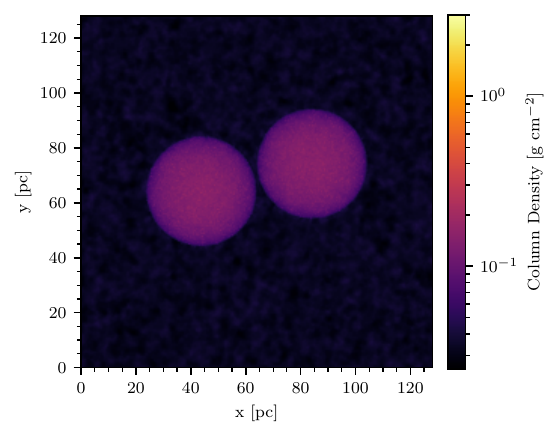}
    \caption{Surface density plot of the initial positions of the clouds.}
    \label{fig:ICS}
\end{figure}

In this paper we adopt the initial conditions provided in W17, which were motivated by observations of GMCs. The simulations are initialised within a (128 pc)$^{3}$ domain of molecular gas with a mean molecular weight $\mu=2.33$ and a helium to hydrogen fraction of 0.1. Two spherical clouds of radius $R_{\text{GMC}}=20$ pc are placed into the domain with their centres separated by $(x,y,z)=(2R_{\text{GMC}},b,0)$ where $b=0.5R_{\text{GMC}}$ (see Figure~\ref{fig:ICS}). Each cloud is initialised with a mass of $M_\text{GMC}=9.3\times10^4 \: {\rm M_{\odot}}$ and an initial temperature of 15~K by randomly distributing 2 million mesh generating points uniformly which constructs our initial cells, achieving an initial mass resolution of 0.0465~M$_\odot$. This results in a hydrogen nucleon density of $n=80.2$~cm$^{-3}$ within the clouds where $n=\frac{\rho}{(1+4A_\mathrm{He})m_\mathrm{p}}$ with $A_\mathrm{He} = 0.1$ being the helium to hydrogen fraction and $m_\mathrm{p}$ being the mass of a proton. Surrounding the clouds is a region of warmer molecular gas with $n=7.14$~cm$^{-3}$ and a temperature of T=150~K. The clouds are not initially in thermal pressure equilibrium with the surrounding material. However, with an initial thermal to gravitational energy ratio of$E_\text{therm}/E_\text{grav}=0.0066$, the clouds are significantly gravitationally bound against thermal pressure, and hence any loss of mass into the low density surrounding medium is small.

A uniform magnetic field of $|B| = 10$ $\umu$G in magnitude is set across the whole domain of the simulation. This results in an Alfv\'en velocity of $v_\text{A}=2.06$~km s$^{-1}$ for each cloud and a mass-to-flux ratio that is 6 times greater than the critical value for both clouds \citep{1974ApJ...193...73G}, making them magnetically supercritical. The angle $\theta$ is the magnetic field inclination from the $x$-axis in the $x$-$y$ plane and is varied between simulations to investigate whether the orientation of the magnetic field affects the star formation rate. (see Table~\ref{tab:IC-table}).

We give the two clouds initial velocities of $v = +v_\text{rel}/2$ and $v = -v_\text{rel}/2$ for the clouds starting at negative and positive $x$, respectively, so that they will later collide at a relative velocity $v = v_{\rm rel}$. We carry out simulations with a range of different $v_{\rm rel}$, so that we can investigate how different collision strengths impact the star formation rate. We also include one case where $v_{\rm rel} = 0$~km s$^{-1}$ (simulation 2), to allow us to investigate what happens in the absence of a collision.\footnote{Note that even in this case, the clouds will eventually collide due to their mutual gravitational attraction. However, this will occur on a timescale $> 10$~Myr, much longer than the period simulated here.} Full details of the velocities used can be found in Table~\ref{tab:IC-table}. Along with the collisional velocity, a turbulent velocity field is also applied to each cloud. The turbulence applied is purely solenoidal and has a 3D velocity dispersion of $\sigma = 3.46$ km s$^{-1}$. This follows a scaling law of $P(k) \propto k^{-4}$. For the velocity field within each cloud the turbulence induced is both supersonic and super-Alfv\'enic, where $\mathcal{M}=\sigma/c_\text{s}=11.6$ and $\mathcal{M}_\text{A} = \sigma/v_\text{A}=1.68 $ respectively. The turbulence induced allows for each cloud to be in virial equilibrium, with the kinematic to gravitational energy ratio for each cloud being $E_\text{kin}/E_\text{grav} = 0.499$.

We also investigate the effect of varying the number of cells per local Jeans length (hereafter referred to as the Jeans refinement parameter, JR). To do this we adopt Jeans refinement as our main cell refinement criterion in {\sc arepo}. Cells are refined by adding additional mesh generating points whenever  $D > \lambda_{\rm J} / {\rm JR}$, where $D$ is the effective diameter of the cell (i.e.\ the diameter of a sphere with the same volume as the mesh cell). We adopt ${\rm JR} = 8$ as our default value for the Jeans parameter, but also investigate the behaviour of runs with ${\rm JR} = 4$ and ${\rm JR} = 16$, as detailed in Table~\ref{tab:IC-table}.

We run the simulations as far as practical to establish a trend in the star formation history. The end time varies due to the adaptive timestep of the simulation, between 3 and 4 Myr. In the presence of close interacting objects, such as binaries, the timestep reduces significantly thus slowing the simulations progression.

\begin{table}
    \centering    
    \caption{Initial conditions that are altered between simulations.}
    \begin{tabular}{lccr}
        \hline
         Simulation & $\theta$ & $v_{\text{rel}}$ & Cells per \\
         & ($^\circ$) & (km s$^{-1}$) & Jeans length\\
         \hline
         1 & 60 & $10$ & 8 \\ 
         2 & 60 & 0 & 8\\
         3 & 60 & $5$ & 8\\
         4 & 60 & $15$ & 8\\
         5 & 0 & $10$ & 8\\
         6 & 30 & $10$ & 8\\
         7 & 90 & $10$ & 8\\
         8 & 60 & $10$ & 4\\
         9 & 60 & $10$ & 16
    \end{tabular}

    \label{tab:IC-table}
\end{table}

\section{Results}
\label{sect:results}

\subsection{Star formation rate}
\label{sect:SFR}

\begin{figure*}
    \centering
    \includegraphics[width=\textwidth]{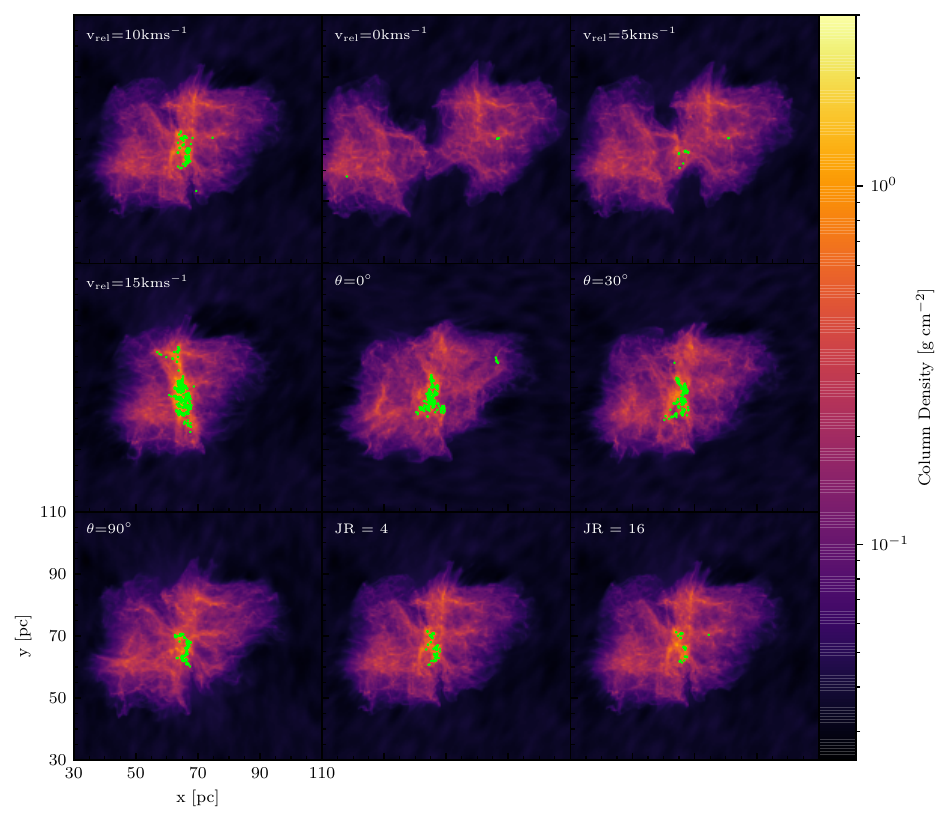}
    \caption{Column density plots of all simulations at $t=2.40$ Myr. The green dots within the plot represent the sink particles that have formed.}
    \label{fig:allsurf}
\end{figure*}

Each simulation results in a different star formation history. For an example, in Figure~\ref{fig:allsurf} it can be seen that the systems evolve differently, creating unique structures. As a result, varying numbers of sink particles form in the simulations, as shown by the green dots in Figure~\ref{fig:allsurf}. To investigate this difference quantitatively, we look at the star formation rate of each of the simulations.

\begin{figure}
    \centering
    \includegraphics[width=\columnwidth]{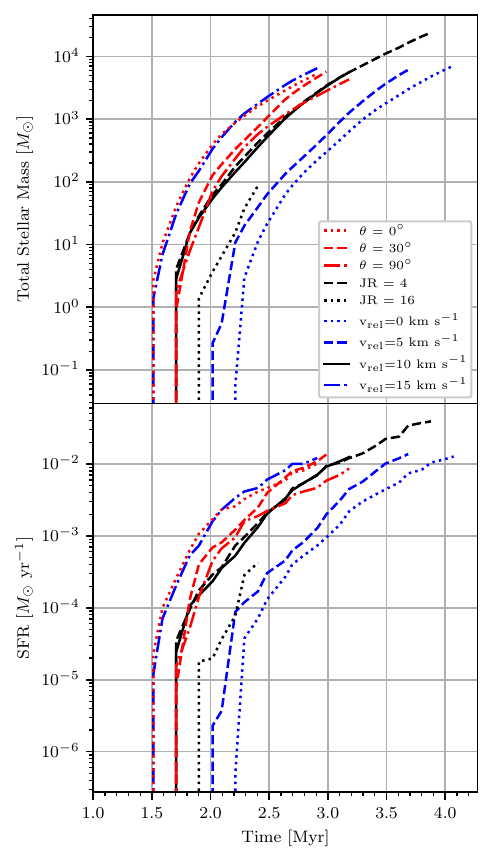}
    \caption{\textit{Top}: Evolution of the mass in sink particles as a function of time in all nine simulations. \textit{Bottom}: Star formation rate as a function of time in all nine simulations.}
    \label{fig:SFR}
\end{figure}

The star formation rate is calculated as:
\begin{equation}
    \label{eq:SFR}
    \Dot{M}_\text{SFR} = \frac{\Delta M}{\Delta t}
\end{equation}
where $\Delta M$ is the difference in the mass of the sinks between the start and end of a time step of length $\Delta t$. These are calculated for each output snapshot available in all nine simulations. The results from these calculations are presented in Figure~\ref{fig:SFR} along with the total mass that has gone into star particles.

Figure~\ref{fig:SFR} shows that there is a clear difference in the time taken for stars to begin forming in the different runs, with a spread in onset times of $\sim 0.7$~Myr between the most extreme cases. Looking in more detail at the results from the individual runs, we see that changing the collisional velocity has the greatest impact on the time required for star formation to begin. Stars start to form sooner in simulations with high collisional velocity than in simulations with low or zero collisional velocity. This behaviour is likely due both to the time it takes for large-scale shocks to form in the colliding clouds, and also to the density enhancement produced by these shocks. Faster collisions produce stronger shocks and hence larger density enhancements.\footnote{Recall that for an isothermal shock, the strength of the density enhancement scales as the square of the Mach number. Although our model GMCs are not isothermal, the equilibrium temperature of the molecular gas varies only weakly with temperature, and so the isothermal result remains a useful guide.} Therefore, higher density star-forming regions with shorter free-fall times are formed with increased collision velocity.

The orientation of the magnetic field has a smaller effect on the time required for star formation to begin. We see a clear difference in behaviour between the case with the magnetic field orientated parallel to the collisional axis, which forms stars after $\sim 1.5$~Myr, and the runs with other magnetic field orientations, in which the onset of star formation is delayed by $\sim0.25$~Myr. The earlier star formation observed when the magnetic field is parallel to the collisional axis is a result of the magnetic field not hindering the compression of the gas along the $x$-axis. This allows the shocked gas to reach the densities required for the first sinks to be created more rapidly compared to the other magnetic field orientations (see also Appendix \ref{sect:mag}). This delayed behaviour agrees well with that of the strong B$_{\rm y}$ simulation of \citet{Dobbs2021}.

Finally, regarding the Jeans refinement variation, a higher resolution tends to delay the onset of star formation. This is likely due to the more highly resolved turbulent velocity field resulting in more disruption of the star-forming gas. This results in the gas being unable to be fully bound, delaying star formation. However, it should be noted that the difference of the onset of star formation between the $JR=8$ and $JR=16$ cases is small, of the order of $\sim 0.2$ Myr, which is about the free-fall time of a dense core.

\begin{figure}
    \centering
    \includegraphics[width=\columnwidth]{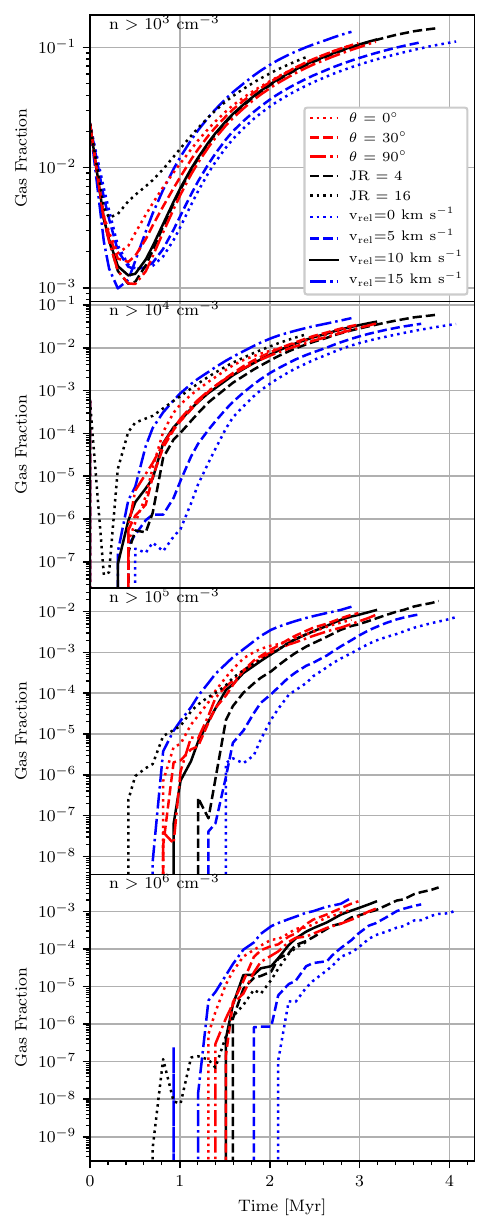}
    \caption{Fraction of gas above the specified density threshold in all nine simulations, plotted as a function of time. The thresholds are indicated in the top-left corner of each panel and are $n> 10^3\: {\rm cm^{-3}}$, $10^4\: {\rm cm^{-3}}$, $10^5\: {\rm cm^{-3}}$ and $10^6\: {\rm cm^{-3}}$ from \textit{top} to \textit{bottom.} We see that increasing the collisional velocity increases the amount of gas above the threshold for all values of the threshold, but that the effect is much stronger for the higher density thresholds.}
    \label{fig:dense_gas}
\end{figure}

It is also informative to look at how the dense gas fraction -- defined here as the fraction of the gas above a density of $10^{4} \: {\rm cm^{-3}}$  with common observational definitions \citep[e.g.][]{Lada2010} -- varies as a function of time in the different simulations. We illustrate this in the bottom three panels of Figure~\ref{fig:dense_gas} and as expected, dense gas is produced more rapidly as we increase the collision velocity, confirming our suspicion that the differences we observe in the timing of the onset of star formation are largely due to the differing amount of time it takes to compress the gas. Changing the orientation of the magnetic field also changes the dense gas fraction, with an appreciably higher fraction produced when the magnetic field is aligned with the axis of the collision. It is also apparent that there is an offset of around 0.5--1.0~Myr between the time taken to produce dense gas at $10^4 \: {\rm cm^{-3}}$ and the onset of star formation. The offset reduces when we consider higher density gas fractions. This offset is easily understood as a consequence of our definition of ``dense'' gas, and the fact that the density threshold for sink particle creation is significantly higher than the value we use in our definition of dense gas. At $n = 10^{4} \: {\rm cm^{-3}}$, the gravitational free-fall time of the gas is $\sim 0.5$~Myr, and so we would expect gas close to this density to require around this long to collapse to stellar densities (in reality) or the sink creation density (in the simulations).\footnote{Note that in a pure free-fall collapse, the time taken to collapse from $10^{4} \: {\rm cm^{-3}}$ to our sink creation density of $\sim 10^{8} \: {\rm cm^{-3}}$ is two orders of magnitude longer than the time taken from collapse from $\sim 10^{8} \: {\rm cm^{-3}}$ to stellar densities.} The offset we see between the appearance of ``dense'' gas and the formation of stars therefore corresponds to the one to two free-fall times required for gas at this density to collapse sufficiently to form stars.\footnote{In the absence of pressure, we would expect the gas to collapse within a single free-fall time. In reality, however, the non-negligible magnetic pressure and kinematics of the gas inevitably delays the collapse (See Sect.~\ref{sect:virial}).
Finally, even once we account for this offset, it is clear that not all of the dense gas that forms in the simulations goes on to form stars. Instead, the star formation efficiency in gas of this density is typically only a few tens of percent.}

From Figure~\ref{fig:dense_gas} we observe significant variation of the gas fractions at early timesteps prior to the steady evolution of the gas density. Most of these variations are transient features in which denser gas forms due to the collision but is dispersed due to pressure gradients briefly forming. For the two lower density thresholds, we also see that there is a small fraction of gas above the threshold present at the beginning of the simulation. This results from the uneven distribution of mesh points within the initial conditions, which results in a certain amount of scatter in the starting densities of the cells.

\begin{figure}
    \centering
    \includegraphics[width=\columnwidth]{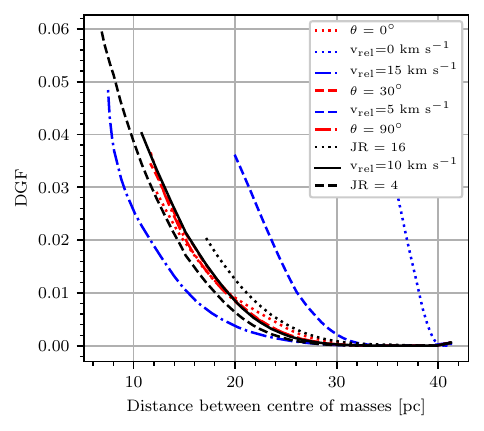}
    \caption{Fraction of dense ($n > 10^{4} \: {\rm cm^{-3}}$) gas in all nine simulations, plotted as a function of the separation between the clouds' centre of masses}.
    \label{fig:dens_sep}
\end{figure}

\begin{figure}
    \centering
    \includegraphics[width=\columnwidth]{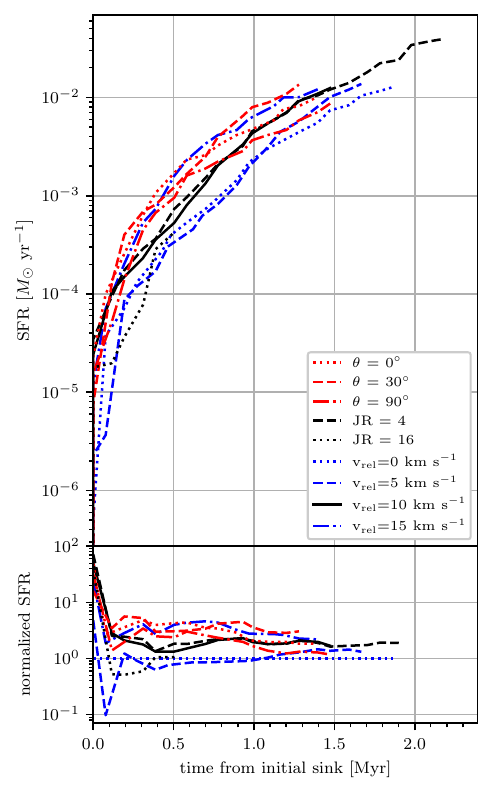}
    \caption{\textit{Top}: Star formation rate varying with time from initial sink formation for all simulations. \textit{Bottom}: Star formation rate normalised to $v_\text{rel}=0$ km s$^{-1}$ star formation rate as a function of time from initial sink formation.}
    \label{fig:sinkSFR}
\end{figure}

The role of the collisions in driving the increase in the dense gas fraction and bringing about the onset of star formation can also be seen clearly if we examine how the dense gas fraction varies as a function of cloud separation, defined as the distance between the centres of mass of the two clouds. We see that most of the simulated clouds follow very similar tracks in this diagram, with substantial quantities of dense gas becoming apparent only once the cloud separation is less than $\sim 25$~pc, i.e.\ once the collision is significantly advanced. The two exceptions are the clouds with $v_{\rm rel} = 0$ and 5~km~s$^{-1}$, which start to develop dense gas while at clearly larger separations, particularly in the $v_{\rm rel} = 0 \: {\rm km \, s^{-1}}$ run. These two runs are also the ones that produce the lowest dense gas fractions at any given time and that form the least number of stars. Our results are therefore consistent with the idea that in these runs, star formation is driven primarily by the collapse of the individual clouds, with the interaction between them playing little role, while in the simulations with higher $v_{\rm rel}$, the collision between the clouds plays a much more substantial role in influencing the star formation rate.

Figure~\ref{fig:sinkSFR} shows that the star formation rate in each simulation follows a similar trend from the point when sinks first form. For a brief period after the first sinks form, the SFR in the runs with $v_{\rm rel} > 0$~km s$^{-1}$ is as much as an order of magnitude larger than in the run with $v_\text{rel} = 0$~km s$^{-1}$, in which the clouds do not collide. However, the difference between the runs quickly decreases, and for the majority of the time covered by our simulations, the SFR in the cloud collision runs is only a factor of 2--3 larger than in the zero velocity run.
This behaviour agrees well with the enhancement in the star formation rate found by \citet{Liow2020} and \citet{2020MNRAS.494..246T} for collision velocities below $20 \: {\rm km \, s^{-1}}$. However, it is substantially smaller than the order of magnitude increase found by W17 despite the similarity between our initial conditions and theirs. To help us better understand the origin of this difference in results, we look in the next section at the virial parameter of the clouds and clumps formed in the simulations. 

\subsection{Impact on the virial parameter}
\label{sect:virial}

In the previous section, we showed that although the cloud collision increases the star formation rate of our colliding clouds compared to the case with no collision, the magnitude of the increase is much smaller than the order of magnitude found by W17. This difference in outcome from their simulations cannot be a consequence of the initial conditions, as we use the same initial conditions as in their study, and hence must be a consequence of the difference in numerical approaches. As we discuss in more detail in Section~\ref{sect:disc}, one of the main differences between our two studies is the algorithm we use to identify star-forming gas. W17 form stars stochastically in gas above a fixed density threshold, with no consideration given to whether or not the gas is gravitationally bound, whereas we use a sink particle based approach in which stars form only in regions that are verifiably bound and collapsing. To understand whether this algorithmic difference can explain the difference in outcome, we explore how the virial parameter of the gas varies on different scales within our set of simulations.

The virial parameter $\eta$ provides an insight into how bound the star-forming region is by comparing the gravitational energy against all other energy contributions. We adapt the definition provided in \citet{2016MNRAS.462.4171B} to include magnetic energy as a factor which counteracts gravitational collapse (Eqn.~\ref{eq:vir}).

\begin{equation}
    \label{eq:vir}
    \eta = \frac{2(E_{\text{kin}}+E_{\text{therm}}+E_{\text{mag}})}{|E_{\text{grav}}|}
\end{equation}

We chose to include the magnetic energy in this calculation as the gas we are considering has a density below the sink formation threshold and does not exist near sinks. It should be noted  that the magnetic energy is not accounted for in the energy check of the sink creation protocol.\footnote{In cases where the magnetic energy is high enough in the sink formation region to prevent collapse, we would also expect it to halt the infall of the gas, i.e.\ independent of the energy check, the region will fail condition (b) of the sink creation protocol and hence will not be converted into a sink.}

In order to make the comparison between the simulations we perform this analysis on the snapshots that are $\sim$1~Myr after the formation of the first sink particle in each simulation. Note that this corresponds to a different physical time in each simulation, but should allow us to compare the clouds when they are at a similar stage in their evolution.

The regions used to determine the virial ratio are chosen by using the local minima of the potential. These are identified as part of the flux calculations within the MHD code and are stored within the outputs. After identification, we then compute the Jeans length for the gas located at the potential minimum. For the purposes of this analysis, we want to avoid potential minima that have already formed stars, so that we can measure the virial ratio accurately without worrying about the confounding effect of the sink particles.\footnote{As sink particles selectively remove the most bound gas, the virial ratio in the remaining gas will inevitably be higher than it would be in the same region if the sink were not present.} We therefore remove from consideration any minima that are located within two Jeans lengths of a sink particle.
We next select all gas cells lying within a single Jeans length of the identified minimum and compute the total gravitational, kinetic, magnetic and thermal energies of this set of gas cells. The gravitational energy, $E_\text{grav}$, is calculated via a direct N-body approach,

\begin{equation}\label{eq:grav_shell}
    E_{\text{grav}} = -\sum_{i=1}^N \sum_{j=i+1}^N \frac{G m_i m_j}{\Delta r_{ij}}
\end{equation}
where $m_i$ and $m_j$ are the masses of the $i$th and $j$th cells, respectively, $\Delta r_{ij}$ is the separation between the cells and $N$ is the total number of cells within the selected region. The kinetic, magnetic and thermal energies, $E_\text{kin}$, $E_\text{mag}$ and $E_\text{therm}$, are calculated as follows:
\begin{eqnarray}
    E_{\text{kin}} & = & \frac{1}{2} \sum_{i=1}^N m_i v_i^2 \,, \label{eq:kin_shell} \\
    E_{\text{mag}} & = & \frac{1}{8\pi} \sum_i |B_i|^2 \Delta V_i \,, \label{eq:Emag} \\
    E_{\text{therm}} & = & \sum_i m_i e_{\text{th}_i} \,, \label{eq:Etherm}
\end{eqnarray}
where $m_i$ is the mass of cell $i$, $v_i$ is the velocity of the cell relative to the potential minimum considered, $|B_i|$ is the magnetic field strength of the cell, $\Delta V_i = m_i/\rho_i$ is the approximate cell volume where $\rho_i$ is the cell density, and $e_{\text{th}_i}$ is the thermal energy per unit mass of the cell.

The number densities associated with the local minima of the potential cover a wide range of values. For the purpose of this analysis we focus on a range of associated hydrogen nuclei number densities from $n = 100$ cm$^{-3}$ to $10^6$ cm$^{-3}$. Our motivation for adopting a lower limit of 100~cm$^{-3}$ is to ensure that are selecting regions associated with the cloud and not the surrounding inter-cloud medium. The upper limit is chosen to match the density threshold for star formation used in W17.
We split the data into four bins of 1 dex in density starting at $10^2$-$10^{3}$ cm$^{-3}$. 

\begin{figure}
    \centering
    \includegraphics[width=\columnwidth]{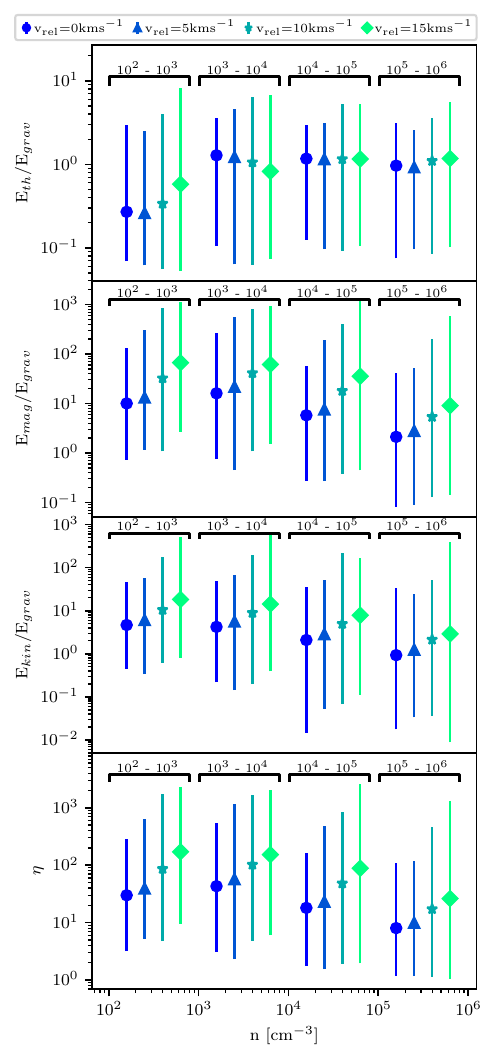}
    \caption{From the top down: The average a) thermal to gravitational energy ratio, b) magnetic to gravitational energy ratio, c) kinetic to gravitational energy ratio, and d) virial parameter for different collisional velocities $v_\text{rel}$ at 1 Myr after first sink formation. The error bars indicate the full range of values obtained. Each density bin is indicated with the corresponding bracket annotation with the separation of the points for ease of reading. The points within a bracket are for the same density bin but are offset in the figure for clarity.} 
    \label{fig:virialpar1}
\end{figure}

In Figure~\ref{fig:virialpar1}, we show the ratio of each of the different components of the energy (thermal, magnetic, kinetic) to the gravitational energy as well as the averaged virial parameter for the hydrogen nuclei number density $n$ bin considered for varying collisional velocity. The error bars indicate the full range of values of each parameter in each bin, i.e.\ the range from the minimum to the maximum value of the parameter.

Upon first inspection we note that the virial parameter is considerably greater than 1 in almost all of the regions we consider. This would imply that they are not gravitationally bound and thus unable to form stars at this point as a result. While this is likely true for some regions, particularly in our lower density bins, it should also be noted that we are considering only a single Jeans length from the potential minimum, and therefore we would expect to recover $\eta \sim 1$ for this region even if the kinetic and magnetic energies were zero.\footnote{We do not recover exactly one because the density and temperature are not constant within the selected region.} Since these components are not zero, it is unsurprising that we recover $\eta \gg 1$ for most of these regions, even the ones that will later form stars. In practice, all that this means is that the radius of the region that will ultimately collapse is greater than a single Jeans length, owing to the additional support provided by the turbulent motions and the magnetic field. Nevertheless, this does provide a clear demonstration of the difficulties involved in identifying star-forming regions based on the properties of individual gas cells, as is the case when applying for example a simple density threshold, since ultimately the answer to the question of whether or not this gas will collapse and form stars depends not only on its own properties but also those of the surrounding gas cells.

For the purposes of the comparison we are making here, we look for differences in the parameters between the simulations considered, and in particular for any general trends. For the ratio of $E_{\rm therm}$ to $E_{\rm grav}$, we do not see a clear trend with collision velocity: the ratio increases with increasing $v_{\rm rel}$ in some density bins, but decreases or remains constant in other density bins. However, we observe a consistent increase in the ratios of $E_{\rm mag}$ to $E_{\rm grav}$ and $E_{\rm kin}$ to $E_{\rm grav}$ with increasing $v_{\rm rel}$ in all density bins. This is then reflected in the behaviour of the virial parameter, which also increases with increasing $v_{\rm rel}$ at all densities. This behaviour is a consequence of an increase in the turbulent motion of the dense molecular gas as the clouds collide with higher velocities. This directly increases the kinetic energy, but also results in additional tangling of the magnetic field, which increases the magnetic field strength and magnetic energy. These motions generally act against gravitational collapse and as a result fewer regions become gravitationally bound and star-forming, leading to a smaller increase in SFR than anticipated by W17.

\begin{figure}
    \centering
    \includegraphics[width=\columnwidth]{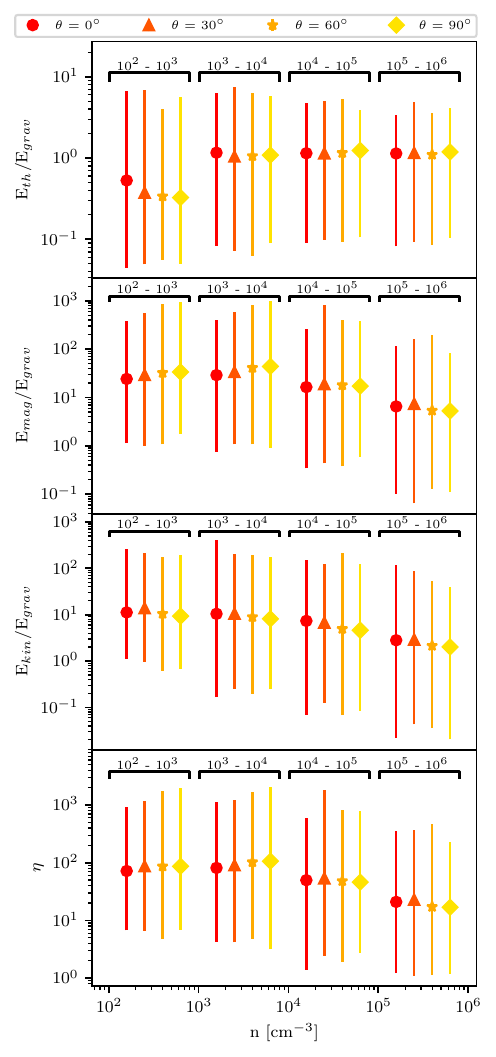}
    \caption{Same as Figure~\ref{fig:virialpar1} but looking at the differences in the inclination of magnetic field}
    \label{fig:virialpar2}
\end{figure}

We have repeated the same analysis for the simulations with differing initial magnetic field inclinations (Fig.~\ref{fig:virialpar2}). Once again, we find that the ratio of $E_{\rm therm}$ to $E_{\rm grav}$ is substantially less than one in our lowest density bin and of order unity in the other bins. In the lowest density bin, we also observe a decrease in the average value of this ratio with increasing inclination angle. A reason for this could be that the gravitational energy calculated for these densities in the $\theta = 0^\circ$ simulation is higher due to compactness. The regions considered for this density bin exist on the outer parts of the clouds and visual inspection of the density maps shows more `flaring' at the edges for greater magnetic field inclinations than for $\theta = 0^\circ$ (see also Figures \ref{fig:mag_abs} and \ref{fig:mag_z}). We see no clear trend in the value of the ratio with magnetic field inclination angle in the other density bins. As for the magnetic and kinetic to gravitational energy ratios, we find the values of these ratios to be similar across the different inclinations within a given density bin. This in turn results in the virial ratio being constant across all magnetic field orientations for each density bin.

\subsection{Flow alignment to magnetic field}
\label{sect:vdotB}

The movement of gas within a magnetic field causes that field to be dragged, distorting the magnetic field in the process. The compression of the field lines causes the flow of gas to slow down as the magnetic field resists the flow. However, this only applies if the direction of flow is not parallel to the magnetic field. The flow remains unhindered if it is parallel to the magnetic field.
Examining the alignment between the flow and the magnetic field can therefore tell us something about the degree to which the field is influencing the motion of the gas.

\begin{figure}
    \centering
    \includegraphics[width=\columnwidth]{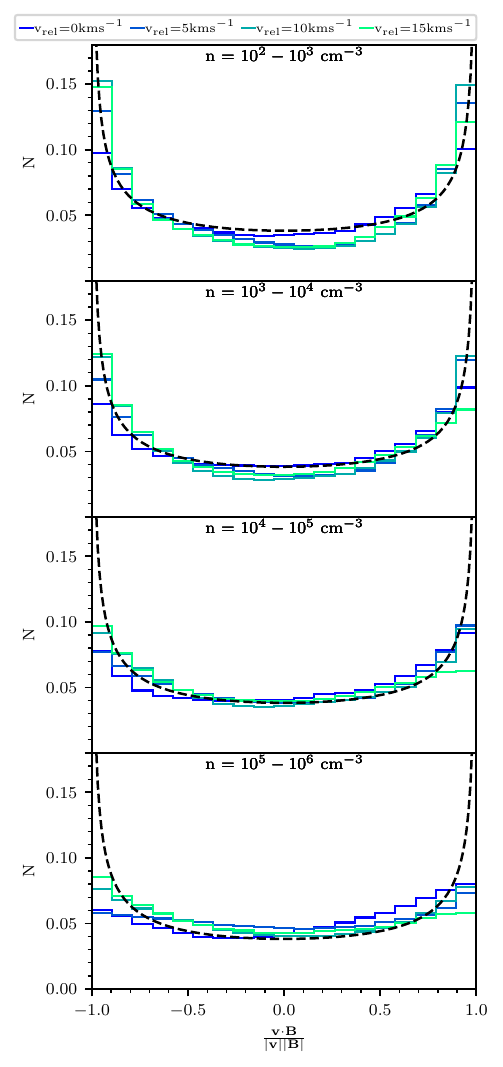}
    \caption{Weighted histogram of the velocity and magnetic field alignment for differing collisional velocities at 1 Myr after first sink formation. Each subplot represents a different density bin. The black dashed line represents the distribution we would expect if the alignment is random.}
    \label{fig:vdotB_vel}
\end{figure}

\begin{figure}
    \centering
    \includegraphics[width=\columnwidth]{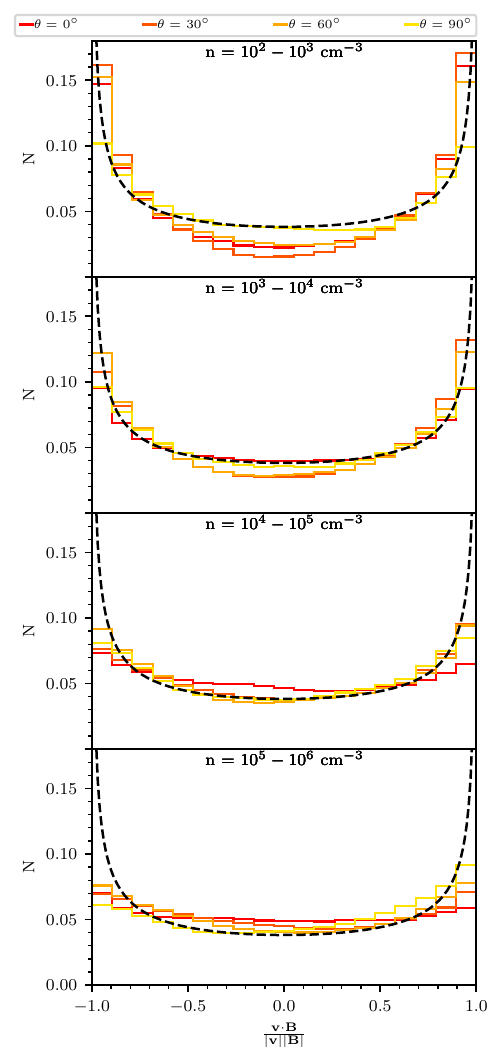}
    \caption{Same as Figure~\ref{fig:vdotB_vel} but for varying magnetic field inclination.}
    \label{fig:vdotB_mag}
\end{figure}

For this purpose, we define the alignment as the dot product of the velocity and magnetic fields normalized by the magnitude of the fields, in other words:\
\begin{equation}
{\rm Alignment} = \frac{\mathbf{v} \cdot \mathbf{B}}{|\mathbf{v}||\mathbf{B}|}
\end{equation} 
This yields the cosine of the angle between the magnetic field and the flow velocity for each Voronoi cell, which has values ranging from $-$1 to 1, with the extremes meaning the fields are parallel whilst a value of 0 means the fields are perpendicular. This corresponds to unrestricted and fully restricted flow by the magnetic fields, respectively. In Figures~\ref{fig:vdotB_vel} \& \ref{fig:vdotB_mag}, we explore how the field alignment varies between our four density bins in runs with different relative velocities and magnetic field inclinations. In each case, the count of the points at each alignment is normalized by the total number of points within
the relevant density bin. The analysis is carried out using the same snapshots as in Section~\ref{sect:virial}.

\begin{figure}
    \centering
    \includegraphics[width=\columnwidth]{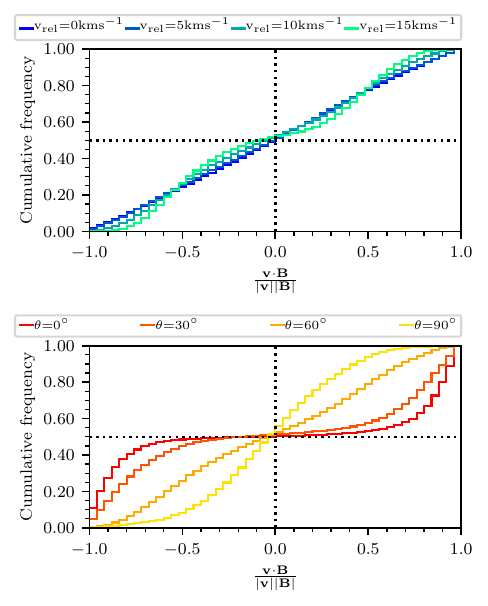}
    \caption{Cumulative distribution of the initial alignment of the simulations at $t=0$ Myr. \textit{Top}: Varying collisional velocity. \textit{Bottom}: Varying magnetic field inclination. The dashed black lines represent where the cumulative distribution would encompass 50\% of the data assuming a Gaussian distribution of alignments. Points in the top-left and bottom-right quadrants are indicative of a skewed distribution.}
    \label{fig:init_vdotb}
\end{figure}

In low density regions ($10^2-10^4$ cm$^{-3}$), we find a distribution of alignments similar to what we would expect for a fully random distribution (indicated in Figures~\ref{fig:vdotB_vel} \& \ref{fig:vdotB_mag} by the black dashed line). In the highest collision velocity runs ($v_\mathrm{rel}\geq10$ km s$^{-1}$), there is a slight skew in the distribution of alignments that we have traced to a similar skewness in the distribution of alignments in the initial conditions (See Fig.~\ref{fig:init_vdotb}). Similarly, in the runs with $\theta = 0^\circ$ and $\theta = 90^\circ$, we also see a clear imprint of the initial conditions, with a preference for parallel alignments in the run with $\theta = 0^\circ$ and for perpendicular alignments in the run with $\theta = 90^\circ$. Overall, there is little indication that the field plays a significant dynamical role in the low density gas.

At higher gas densities ($10^4-10^6$ cm$^{-3}$), we no longer see any clear imprint of the initial conditions. The distribution of alignments close to 0.0 becomes very similar to what we would expect for a fully random distribution. However, there is a clear deficit of alignments close to $-$1 and $+$1 that becomes more pronounced as the density increases. In other words, in dense gas, the gas flow is predominantly perpendicular to the field, rather than parallel to it. This behaviour is consistent with our expectations for magnetically supercritical gas: the flow of gas perpendicular to the magnetic field compresses and strengthens the field, while the flow of gas along the field lines leaves the field strength unaltered, resulting in a field alignment that becomes increasingly perpendicular as the compression continues. This behaviour has been seen in other simulations \citep[such as][]{Soler2017} and is a result of velocity field convergence, $\nabla\cdot v<0$ \citep{Boldyrev2006,Matthaeus2008}.

One further important take away from these results is that in the densest gas, there are only minor differences in the field alignment between the different runs, consistent with a picture in which the small-scale behaviour of the field in dense regions is primarily determined by the local velocity field and not by the large-scale details of the collision.

\section{Discussion}
\label{sect:disc}
\subsection{Comparison with W17}
Our simulations demonstrate that rapid collisions between clouds lead to an earlier onset of star formation than in clouds that collide slowly or not at all. However, the way in which star formation proceeds once the process has set in is remarkably similar in all of our simulations. Note we find a persistent enhancement of a factor of 2--3 in the star formation rate in our colliding clouds compared to our $v_\mathrm{rel} = 0$~km s$^{-1}$ control run. The earlier onset of star formation and the increase in the  star formation rate due to the collision are similar to the results reported by W17 for the same initial conditions. Contrary to W17, we find a only a very weak dependence of the SFR on the collisional velocity, $v_\text{rel}$.

It is worthwhile considering possible methodological reasons for this difference. Aside from our use of a different magnetohydrodynamical code (\citeauthor{2017ApJ...841...88W} use {\sc Enzo}, we use {\sc Arepo}), our simulations differ from those of W17 in two main respects: our star formation algorithm and our treatment of chemistry and cooling.

\subsubsection{Star formation protocol}\label{sect:SFdiff}
In our simulations we make use of sink particles as our representation of stars/protostellar systems, which are able to continue accreting after forming (see Sect~\ref{sect:SF}). In contrast, W17 make use of star particles that are formed stochastically with a fixed efficiency per free-fall time in gas cells on the finest level of refinement in their simulation that satisfy a suitably chosen criterion. In their ``density-regulated'' models, this criterion is a simple density threshold: star formation is permitted only in gas denser than $n_{\rm th} = 10^{6} \: {\rm cm^{-3}}$. In their ``magnetically-regulated'' models, on the other hand, star formation is permitted only in cells that are magnetically supercritical, although their adoption of a fixed star particle mass also acts as an effective density threshold, preventing stars from forming in cells less dense than $3.55 \times 10^{5} \: {\rm cm^{-3}}$ in most of their ``magnetically-regulated'' runs. Notably, W17 do not require the gas flow to be converging or the gas to be gravitationally bound in order for it to be eligible to form stars. 

A final difference between the treatment of star formation in the two approaches is that the star particles formed in the W17 simulations have a fixed mass from the moment that they form, whereas our sink particles can continue to accrete mass as they age. W17 argue that fixing the star particle mass is a way of approximately accounting for the effects of stellar feedback, but 
a comparison between our runs and those of W17 shows that we actually recover much lower star formation rates, by around an order of magnitude, even without this restriction. Similar to the studies by W17, we also do not take stellar feedback into account in this set of simulations. Neglecting supernovae is justified because our computation covers a period of only a few  Myr after the onset of star formation (see Figure \ref{fig:sinkSFR}), which is shorter than the time required for the first supernovae to occur \citep{Kippenhahn12}. Ignoring stellar winds and radiation is also justified at early times, but is less valid once the total stellar mass exceeds a few hundred solar masses, as at this point we would expect to have formed at least a few stars massive enough to start ionising their surroundings and driving strong stellar winds. Nevertheless, the importance of these forms of feedback for gas removal depends very much on the properties of the star cluster and its parental cloud \cite[e.g.][]{Rahner17, Haid18, Rahner19}, and it is unclear how much impact they would have in our simulated clouds.

We also note that our main results do not depend on this simplification. The difference between the star formation rates in the different runs becomes apparent very early on, long before stellar feedback from massive stars could possibly play a role in the evolution of the clouds, and so 
even if we were to have terminated our simulations at the point that the first massive star forms in each case, we would have come to the same conclusions. Finally, the neglect of feedback from protostellar outflows likely does affect the star formation rate at all times in the simulation. However, previous numerical studies have shown that the impact of this form of feedback is always negative: it reduces the star formation rate by a factor of around 2--3 compared to models that do not account for it \citep[see e.g.][]{Federrath2015,Hu2022}. Its absence from our simulations therefore cannot explain the discrepancy between our results and those of W17.

We have assumed that the impact of stellar feedback on the star formation rate is always negative. This is a good approximation on the scales of individual star-forming clouds \citep{Grisdale2017}, but on larger scales it is possible that feedback could in some circumstances be positive, triggering the compression of gas and the onset of star formation somewhere else in the galaxy  \citep{Shore1981}. However, exploring the effect of feedback on these scales lies far outside of the scope of this paper.

The simplest explanation for the difference between our study and the results of W17 is the different star formation criteria applied in the two approaches. W17 motivate their adoption of a simple density or mass-to-flux ratio threshold on the grounds of the limited resolution of their simulations (smallest cell size $\Delta x  = 0.125$~pc), which does not make them confident that they can resolve the small scale structure of the gas well enough to apply a more complicated criterion. Our spatial resolution is much better in gravitationally collapsing regions, thanks to the Jeans refinement criterion, with our minimum cell size becoming as small as $\sim10^{-4}$~pc shortly before sink particle formation. We therefore do a much better job of resolving the structure of the dense gas and can easily distinguish between dense gas that is gravitationally bound and star-forming and dense gas that is not gravitationally bound. Since there is evidence that a significant fraction of the dense gas produced in the collisions is not gravitationally bound, as we have already seen in Section~\ref{sect:virial}, it is therefore unsurprising that we recover much smaller star formation rates than in the W17 simulations, but that are much more in line with the results found by other studies using a sink particle based approach \citep{2020MNRAS.494..246T,DobbsLiowRieder2020}. 

The sink particle studies mentioned here do differ in their sink protocols. All studies include a sink creation density threshold and require the sink-forming cell (or particle) to lie at a potential minimum. Whilst this study and \citet{Liow2020} have a similar protocol (this study having a higher sink creation density threshold), \citet{2020MNRAS.494..246T} do not have either a gas convergence or a boundedness check in their sink creation protocol. Instead, they opt for further constraints on where sinks form that required that the sink-forming gas is not tidally interacting with other sinks, and that it is capable of undergoing free-fall collapse before interacting with another sink.

The lower density threshold of \citet{Liow2020} could conceivably result in more sink particles forming than in our simulation, but the factor by which the SFR is enhanced should be unaffected provided the threshold density is consistent in simulations with and without collision. The same argument can be made for \citet{2020MNRAS.494..246T}: 
 we would again expect their comparison of star formation rates between colliding and non-colliding runs to be meaningful, since their protocol is consistent across the simulations.

The differences in the sink creation protocol has scope to yield significant differences in the SFR enhancement across the studies, yet it has not. This suggests that the active accretion onto sink particles limits the enhancement in the SFR with GMC collisions in comparison to fixed mass star particles.

That all said, we stress that this remains merely our most plausible hypothesis for the difference in outcomes. To confirm this would require us to carry out simulations similar to the ones presented here but using exactly the same star formation prescription as in W17 as well as simulations with the same initial conditions as \citet{2020MNRAS.494..246T} and \citet{Liow2020}, a task which is outside of the scope of our current study.

It should also be noted that the use of star particles in the context of molecular cloud/GMC-scale star formation is uncommon. Previous works looking at colliding clouds or flows mostly use a sink particle based prescription for star formation (for example \citealt{2020MNRAS.494..246T,DobbsLiowRieder2020,Dobbs2021}). A star particle approach is more commonly used in galaxy-scale and larger simulations where individual GMCs are at best barely resolved. On these scales the justification for a probabilistic approach to star formation is acceptable as the gravitationally-bound cores within GMCs are not resolved and the GMCs themselves are not necessarily gravitationally bound \citep[e.g., see][]{Dobbs2011b}.

\subsubsection{Treatment of chemistry and cooling}\label{sect:chemdiff}
Although we consider the difference in the star formation algorithm to be the most likely cause of the difference between our results and those of W17, it is worthwhile examining whether the difference in our chemical and thermal treatment may also play a role here. As previously mentioned, in our simulations, we use a modified version of the \citet{2017ApJ...843...38G} chemical network, together with the atomic and molecular cooling function described in \citet{2019MNRAS.486.4622C}.
On the other hand, W17 make use of the {\sc Grackle} library \citep{2017MNRAS.466.2217S}. The chemical networks provided in {\sc Grackle} allow one to model the non-equilibrium chemistry of hydrogen (including H$_{2}$ formation and destruction), but do not account for the chemistry of metals such as carbon or oxygen. To account for heating and cooling due to metals, {\sc Grackle} uses a table-based approach: the relevant rates are interpolated from a set of tables in which the rates are given as a function of density, temperature and (optionally) metallicity. {\sc Grackle} includes several sets of tables computed using the {\sc Cloudy} photodissociation region (PDR) code \citep{1998PASP..110..761F}, but also allows the user to supply their own. 
W17 make use of this latter option, using a set of cooling tables generated by \citet{2015ApJ...811...56W} using the PyPDR code \citep{2019ascl.soft05027B}.

\begin{figure}
    \centering
    \includegraphics[width=\columnwidth]{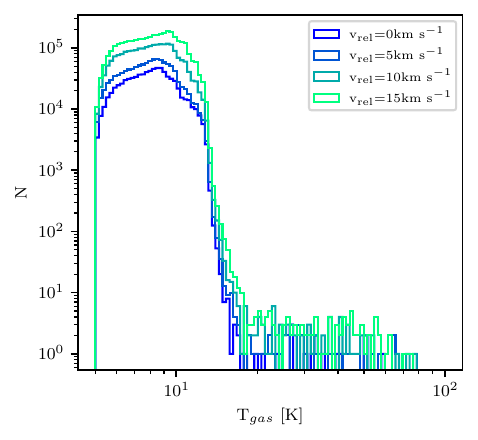}
    \caption{Temperature distribution of all gas cells of $n\geq10^{5}$cm$^{-3}$ for different collisional velocities at 1 Myr after initial sink particle creation.}
    \label{fig:temp}
\end{figure}

\begin{figure*}
    \centering
    \includegraphics[width=\textwidth]{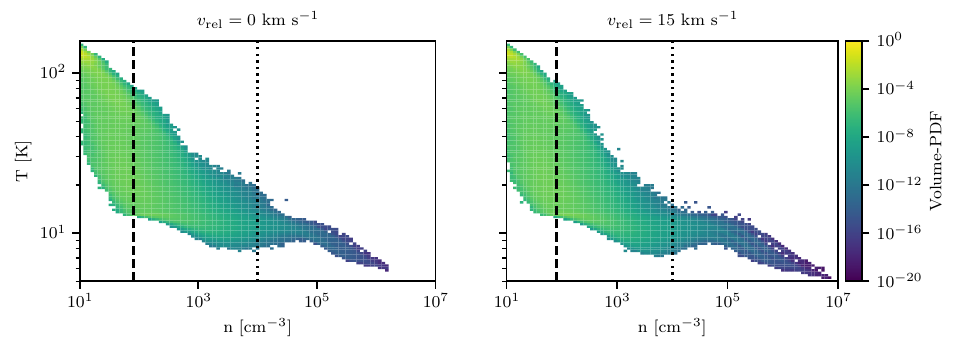}
    \caption{Volume-weighted temperature-density phase diagram for $v_{\rm rel} = 0 \: {\rm km \, s^{-1}}$ (left panel) and $v_{\rm rel} = 15 \: {\rm km \, s^{-1}}$ (right panel), shown for the snapshot immediately prior to the formation of the first sink particle. The dashed vertical line is the initial density of the clouds and the dotted vertical line is the density above which we consider the gas to be dense. The impact of $v_{\rm rel}$ on the temperature distribution is small, owing to the short cooling time of the gas.}
    \label{fig:temp_phase}
\end{figure*}

The impact of this difference in approach can be seen if we compare the temperature distribution of the dense gas in our simulations (shown in Figure~\ref{fig:temp}) with the temperature distribution of the star-forming gas in the W17 simulations (shown in their figure 6). We find that most of the dense gas in our simulations has a temperature in the range $5 < T < 20$~K, with a small tail in the distribution extending up to $\sim 100$~K. We also find that there is no clear difference in the temperature distribution of the dense gas between the colliding ($v_\mathrm{rel}>0$~km s$^{-1}$) and the stationary ($v_\mathrm{rel}=0$~km s$^{-1}$) clouds. On the other hand, W17 recover a somewhat broader temperature distribution, covering the range $10 < T < 40$~K in their non-colliding runs and extending up to $\sim 100$~K in the colliding runs. This broader temperature distribution may be a consequence of the stronger UV field adopted in their calculation -- they assume $G_{0} = 4$, compared to $G_{0} = 1.7$ here -- or may be due to some other aspect of the way in which cooling is treated in the two sets of simulations. Whatever the reason, it is clear that the difference in temperature distributions cannot explain the different star formation rates recovered in the simulations: the dense gas in our simulations is colder on average than that that in the W17 models and hence has less thermal support, meaning that it should be more likely to form dense regions and then stars, rather than less likely. Figure~\ref{fig:temp_phase} demonstrates this point by showing the temperature-density distribution of the gas in the simulations with $v_\mathrm{rel}=0$~km s$^{-1}$ and $v_\mathrm{rel}=15$~km s$^{-1}$ at the snapshot just before sink formation. We see that the phase diagram is similar in both simulations, although there is more dense gas in the run with $v_\mathrm{rel}=15$~km s$^{-1}$. If the difference between the simulations were due solely to the difference in the temperature distributions, we would expect to find a higher SFR in our simulations than in W17, which is the opposite of what we actually see.

\subsection{Do cloud collisions trigger star formation?}
There is growing observational evidence, summarized recently by \citet{2021PASJ...73S...1F}, that the formation of massive stars is often associated with molecular clouds that show signs of having undergone a cloud-cloud collision. This suggests that collisions are generally the cause of the formation of the massive stars, i.e.\ that they would not have formed in the absence of the collision. Unfortunately, observational studies of this issue have an obvious drawback: they can only tell us what did happen, not what would have happened had the situation been different. Simulations, on the other hand, allow us to directly compare the outcome with or without a collision, or with different parameters for the collision, enabling us to better understand the extent to which the collision actually triggers star formation.

In the work presented here, we have insufficient dynamical range to follow the formation of individual stars, and so we cannot directly address the question of whether the collision makes massive star formation more likely. However, we can explore the more general question of whether star formation overall is triggered by the collision between our simulated clouds. Here, the lesson of the simulations is somewhat mixed. It is clear from the fact that star formation occurs in our $v_\mathrm{rel} = 0$~km s$^{-1}$ control run that the cloud collision is not required in order for the clouds to begin forming stars, i.e.\ the collision does not trigger star formation in the sense that the clouds would otherwise remain starless. That said, the fact that star formation begins earlier in the colliding clouds and proceeds at a slightly higher rate are both indicative of the collision having a positive effect overall on the star formation efficiency of the cloud. It is possible that this enhancement of star formation would eventually be lost if we were to simulate the clouds for a much longer period. Conversely, it is also possible that the difference in efficiency would persist, particularly if the cloud lifetime is short. Ultimately, resolving this will require simulations that follow the evolution of the clouds for much longer periods that cover the long-scale collapse of the cloud, beyond the  initial collision, and that also accounts for the stellar feedback processes responsible for dispersing them. However, this is outside of the scope of our current study.

\section{Conclusions}
\label{sect:concl}

In this paper we presented the results of a series of simulations of the collision of two magnetised molecular clouds with mean hydrogen nuclei number densities $n \sim 80 \: {\rm cm^{-3}}$ embedded in a warm, diffuse intercloud medium. In our simulations, we varied the relative velocity of the clouds, the inclination of the magnetic field relative to the collision axis, and the level of Jeans refinement adopted, and investigated the impact that these variations have on the resulting star formation rate. We found that that the different conditions caused star formation to occur at different times but that once star formation had begun, the subsequent evolution of the star formation rate was very similar in all of the simulations. Colliding clouds appear to form stars at a faster rate than clouds that do not collide, suggestive of some degree of triggering of star formation, but the difference in the star formation rates is around a factor of two to three, in line with the results reported by \citet{2020MNRAS.494..246T} but much smaller than the order of magnitude increase found by \citet{2017ApJ...841...88W}. 

We further investigated the virial parameter of {regions of gas around potential minima in our simulations, and how these virial parameters depend on the collisional velocity between the clouds. We found that the virial parameters in the potential minima were higher for the higher collisonal velocities, especially in the high-density, post-shocked gas, with higher amounts of turbulence in the dense gas. Although the higher collision velocity simulations are found to create more high denisty gas, our analysis demonstrates that much of this gas is not gravitationally bound, which explains why the correlation between collision velocity and star formation rate is so weak.}

\section*{Acknowledgements}

GHH would like to thank Elizabeth Watkins, Thomas Williams and Gerwyn Jones for their continuous help throughout this work. We would also like to thank Clare Dobbs and the anonymous referees for their useful comments. GHH and PCC acknowledge support from the StarFormMapper project, funded by the European Union's Horizon 2020 research and innovation programme under grant agreement No 687528. GHH, SCOG and RSK are thankful for the support from the Deutsche Forschungsgemeinschaft (DFG) via the Collaborative Research Centre (SFB 881, Project-ID 138713538) ``The Milky Way System'' (sub-projects A1, B1, B2 and B8) and from the Heidelberg Cluster of Excellence (EXC 2181 - 390900948), ``STRUCTURES: A unifying approach to emergent phenomena in the physical world, mathematics, and complex data'', funded by the German Excellence Strategy. SCOG and RSK also acknowledge funding form the European Research Council in the ERC Synergy Grant ``ECOGAL -- Understanding our Galactic ecosystem: From the disk of the Milky Way to the formation sites of stars and planets'' (project ID 855130). The team benefitted from computing resources provided by the State of Baden-W\"urttemberg through bwHPC and DFG through grant INST 35/1134-1 FUGG, and from the data storage facility SDS@hd supported through grant INST 35/1314-1 FUGG. We also thank for computing time provided by the Leibniz Computing Centre (LRZ) for project pr74nu. 

\textbf{\textit{Software}:} \textsc{AREPO} \citep{2010MNRAS.401..791S}, \textsc{NumPy} \citep{harris2020array}, \textsc{Matplotlib} \citep{Hunter:2007}, \textsc{Astropy} \citep{astropy:2013,astropy:2018}

\section*{Data availability}
The data underlying this article will be shared on reasonable request to the corresponding author.

%%%%%%%%%%%%%%%%%%%%%%%%%%%%%%%%%%%%%%%%%%%%%%%%%%



\bibliographystyle{mnras}
\bibliography{ref} % if your bibtex file is called example.bib

\begin{thebibliography}{}
\makeatletter
\relax
\def\mn@urlcharsother{\let\do\@makeother \do\$\do\&\do\#\do\^\do\_\do\%\do\~}
\def\mn@doi{\begingroup\mn@urlcharsother \@ifnextchar [ {\mn@doi@}
  {\mn@doi@[]}}
\def\mn@doi@[#1]#2{\def\@tempa{#1}\ifx\@tempa\@empty \href
  {http://dx.doi.org/#2} {doi:#2}\else \href {http://dx.doi.org/#2} {#1}\fi
  \endgroup}
\def\mn@eprint#1#2{\mn@eprint@#1:#2::\@nil}
\def\mn@eprint@arXiv#1{\href {http://arxiv.org/abs/#1} {{\tt arXiv:#1}}}
\def\mn@eprint@dblp#1{\href {http://dblp.uni-trier.de/rec/bibtex/#1.xml}
  {dblp:#1}}
\def\mn@eprint@#1:#2:#3:#4\@nil{\def\@tempa {#1}\def\@tempb {#2}\def\@tempc
  {#3}\ifx \@tempc \@empty \let \@tempc \@tempb \let \@tempb \@tempa \fi \ifx
  \@tempb \@empty \def\@tempb {arXiv}\fi \@ifundefined
  {mn@eprint@\@tempb}{\@tempb:\@tempc}{\expandafter \expandafter \csname
  mn@eprint@\@tempb\endcsname \expandafter{\@tempc}}}

\bibitem[\protect\citeauthoryear{{Andr{\'e}} et~al.,}{{Andr{\'e}}
  et~al.}{2010}]{2010A&A...518L.102A}
{Andr{\'e}} P.,  et~al., 2010, \mn@doi [\aap] {10.1051/0004-6361/201014666},
  \href {https://ui.adsabs.harvard.edu/abs/2010A&A...518L.102A} {518, L102}

\bibitem[\protect\citeauthoryear{{Anicich} \& {Huntress}}{{Anicich} \&
  {Huntress}}{1986}]{1986ApJS...62..553A}
{Anicich} V.~G.,  {Huntress} W.~T. J.,  1986, \mn@doi [\apjs] {10.1086/191151},
  \href {https://ui.adsabs.harvard.edu/abs/1986ApJS...62..553A} {62, 553}

\bibitem[\protect\citeauthoryear{{Astropy Collaboration} et~al.,}{{Astropy
  Collaboration} et~al.}{2013}]{astropy:2013}
{Astropy Collaboration} et~al., 2013, \mn@doi [\aap]
  {10.1051/0004-6361/201322068}, \href
  {http://adsabs.harvard.edu/abs/2013A\%26A...558A..33A} {558, A33}

\bibitem[\protect\citeauthoryear{{Astropy Collaboration} et~al.,}{{Astropy
  Collaboration} et~al.}{2018}]{astropy:2018}
{Astropy Collaboration} et~al., 2018, \mn@doi [\aj] {10.3847/1538-3881/aabc4f},
  \href {https://ui.adsabs.harvard.edu/abs/2018AJ....156..123A} {156, 123}

\bibitem[\protect\citeauthoryear{{Badnell}}{{Badnell}}{2006}]{2006ApJS..167..334B}
{Badnell} N.~R.,  2006, \mn@doi [\apjs] {10.1086/508465}, \href
  {https://ui.adsabs.harvard.edu/abs/2006ApJS..167..334B} {167, 334}

\bibitem[\protect\citeauthoryear{{Badnell} et~al.,}{{Badnell}
  et~al.}{2003}]{2003A&A...406.1151B}
{Badnell} N.~R.,  et~al., 2003, \mn@doi [\aap] {10.1051/0004-6361:20030816},
  \href {https://ui.adsabs.harvard.edu/abs/2003A&A...406.1151B} {406, 1151}

\bibitem[\protect\citeauthoryear{{Balfour}, {Whitworth}, {Hubber}  \&
  {Jaffa}}{{Balfour} et~al.}{2015}]{Balfour2015}
{Balfour} S.~K.,  {Whitworth} A.~P.,  {Hubber} D.~A.,   {Jaffa} S.~E.,  2015,
  \mn@doi [\mnras] {10.1093/mnras/stv1772}, \href
  {https://ui.adsabs.harvard.edu/abs/2015MNRAS.453.2471B} {453, 2471}

\bibitem[\protect\citeauthoryear{{Barlow}}{{Barlow}}{1984}]{1984PhDT.......142B}
{Barlow} S.~E.,  1984, PhD thesis, UNIVERSITY OF COLORADO AT BOULDER.

\bibitem[\protect\citeauthoryear{{Bate}, {Bonnell}  \& {Price}}{{Bate}
  et~al.}{1995}]{1995MNRAS.277..362B}
{Bate} M.~R.,  {Bonnell} I.~A.,   {Price} N.~M.,  1995, \mn@doi [\mnras]
  {10.1093/mnras/277.2.362}, \href
  {https://ui.adsabs.harvard.edu/abs/1995MNRAS.277..362B} {277, 362}

\bibitem[\protect\citeauthoryear{{Bertelli Motta}, {Clark}, {Glover}, {Klessen}
   \& {Pasquali}}{{Bertelli Motta} et~al.}{2016}]{2016MNRAS.462.4171B}
{Bertelli Motta} C.,  {Clark} P.~C.,  {Glover} S.~C.~O.,  {Klessen} R.~S.,
  {Pasquali} A.,  2016, \mn@doi [\mnras] {10.1093/mnras/stw1921}, \href
  {https://ui.adsabs.harvard.edu/abs/2016MNRAS.462.4171B} {462, 4171}

\bibitem[\protect\citeauthoryear{{Blitz}}{{Blitz}}{1993}]{Blitz1993}
{Blitz} L.,  1993, in {Levy} E.~H.,  {Lunine} J.~I.,  eds, Protostars and
  Planets III. p.~125

\bibitem[\protect\citeauthoryear{{Boldyrev}}{{Boldyrev}}{2006}]{Boldyrev2006}
{Boldyrev} S.,  2006, \mn@doi [\prl] {10.1103/PhysRevLett.96.115002}, \href
  {https://ui.adsabs.harvard.edu/abs/2006PhRvL..96k5002B} {96, 115002}

\bibitem[\protect\citeauthoryear{{Bruderer}}{{Bruderer}}{2019}]{2019ascl.soft05027B}
{Bruderer} S.,  2019, {PyPDR: Python Photo Dissociation Regions} (\mn@eprint
  {ascl} {1905.027})

\bibitem[\protect\citeauthoryear{Cabral \& Leedom}{Cabral \&
  Leedom}{1993}]{10.1145/166117.166151}
Cabral B.,  Leedom L.~C.,  1993, in Proceedings of the 20th Annual Conference
  on Computer Graphics and Interactive Techniques. SIGGRAPH '93.
Association for Computing Machinery, New York, NY, USA, pp 263--270,
  \mn@doi{10.1145/166117.166151}, \url {https://doi.org/10.1145/166117.166151}

\bibitem[\protect\citeauthoryear{{Carty}, {Goddard}, {K{\"o}hler}, {Sims}  \&
  {Smith}}{{Carty} et~al.}{2006}]{2006JPCA..110.3101C}
{Carty} D.,  {Goddard} A.,  {K{\"o}hler} S. P.~K.,  {Sims} I.~R.,   {Smith} I.
  W.~M.,  2006, \mn@doi [Journal of Physical Chemistry A] {10.1021/jp054429u},
  \href {https://ui.adsabs.harvard.edu/abs/2006JPCA..110.3101C} {110, 3101}

\bibitem[\protect\citeauthoryear{{Chevance} et~al.,}{{Chevance}
  et~al.}{2020}]{Chevance2020}
{Chevance} M.,  et~al., 2020, \mn@doi [\mnras] {10.1093/mnras/stz3525}, \href
  {https://ui.adsabs.harvard.edu/abs/2020MNRAS.493.2872C} {493, 2872}

\bibitem[\protect\citeauthoryear{{Chira}, {Beuther}, {Linz}, {Schuller},
  {Walmsley}, {Menten}  \& {Bronfman}}{{Chira}
  et~al.}{2013}]{2013A&A...552A..40C}
{Chira} R.~A.,  {Beuther} H.,  {Linz} H.,  {Schuller} F.,  {Walmsley} C.~M.,
  {Menten} K.~M.,   {Bronfman} L.,  2013, \mn@doi [\aap]
  {10.1051/0004-6361/201219567}, \href
  {https://ui.adsabs.harvard.edu/abs/2013A&A...552A..40C} {552, A40}

\bibitem[\protect\citeauthoryear{{Clark}, {Glover}  \& {Klessen}}{{Clark}
  et~al.}{2012}]{2012MNRAS.420..745C}
{Clark} P.~C.,  {Glover} S. C.~O.,   {Klessen} R.~S.,  2012, \mn@doi [\mnras]
  {10.1111/j.1365-2966.2011.20087.x}, \href
  {https://ui.adsabs.harvard.edu/abs/2012MNRAS.420..745C} {420, 745}

\bibitem[\protect\citeauthoryear{{Clark}, {Glover}, {Ragan}  \&
  {Duarte-Cabral}}{{Clark} et~al.}{2019}]{2019MNRAS.486.4622C}
{Clark} P.~C.,  {Glover} S. C.~O.,  {Ragan} S.~E.,   {Duarte-Cabral} A.,  2019,
  \mn@doi [\mnras] {10.1093/mnras/stz1119}, \href
  {https://ui.adsabs.harvard.edu/abs/2019MNRAS.486.4622C} {486, 4622}

\bibitem[\protect\citeauthoryear{{Dame}, {Hartmann}  \& {Thaddeus}}{{Dame}
  et~al.}{2001}]{2001ApJ...547..792D}
{Dame} T.~M.,  {Hartmann} D.,   {Thaddeus} P.,  2001, \mn@doi [\apj]
  {10.1086/318388}, \href
  {https://ui.adsabs.harvard.edu/abs/2001ApJ...547..792D} {547, 792}

\bibitem[\protect\citeauthoryear{Dedner, Kemm, Kr\"{o}ner, Munz, Schnitzer  \&
  Wesenberg}{Dedner et~al.}{2002}]{DEDNER2002645}
Dedner A.,  Kemm F.,  Kr\"{o}ner D.,  Munz C.-D.,  Schnitzer T.,   Wesenberg
  M.,  2002, \mn@doi [Journal of Computational Physics]
  {10.1006/jcph.2001.6961}, 175, 645

\bibitem[\protect\citeauthoryear{{Dobbs} \& {Wurster}}{{Dobbs} \&
  {Wurster}}{2021}]{Dobbs2021}
{Dobbs} C.~L.,  {Wurster} J.,  2021, \mn@doi [\mnras] {10.1093/mnras/stab150},
  \href {https://ui.adsabs.harvard.edu/abs/2021MNRAS.502.2285D} {502, 2285}

\bibitem[\protect\citeauthoryear{{Dobbs}, {Burkert}  \& {Pringle}}{{Dobbs}
  et~al.}{2011}]{Dobbs2011b}
{Dobbs} C.~L.,  {Burkert} A.,   {Pringle} J.~E.,  2011, \mn@doi [\mnras]
  {10.1111/j.1365-2966.2011.18371.x}, \href
  {https://ui.adsabs.harvard.edu/abs/2011MNRAS.413.2935D} {413, 2935}

\bibitem[\protect\citeauthoryear{{Dobbs}, {Pringle}  \&
  {Duarte-Cabral}}{{Dobbs} et~al.}{2015}]{2015MNRAS.446.3608D}
{Dobbs} C.~L.,  {Pringle} J.~E.,   {Duarte-Cabral} A.,  2015, \mn@doi [\mnras]
  {10.1093/mnras/stu2319}, \href
  {https://ui.adsabs.harvard.edu/abs/2015MNRAS.446.3608D} {446, 3608}

\bibitem[\protect\citeauthoryear{{Dobbs}, {Liow}  \& {Rieder}}{{Dobbs}
  et~al.}{2020}]{DobbsLiowRieder2020}
{Dobbs} C.~L.,  {Liow} K.~Y.,   {Rieder} S.,  2020, \mn@doi [\mnras]
  {10.1093/mnrasl/slaa072}, \href
  {https://ui.adsabs.harvard.edu/abs/2020MNRAS.496L...1D} {496, L1}

\bibitem[\protect\citeauthoryear{{Draine}}{{Draine}}{1978}]{1978ApJS...36..595D}
{Draine} B.~T.,  1978, \mn@doi [\apjs] {10.1086/190513}, \href
  {https://ui.adsabs.harvard.edu/abs/1978ApJS...36..595D} {36, 595}

\bibitem[\protect\citeauthoryear{{Federrath}}{{Federrath}}{2015}]{Federrath2015}
{Federrath} C.,  2015, \mn@doi [\mnras] {10.1093/mnras/stv941}, \href
  {https://ui.adsabs.harvard.edu/abs/2015MNRAS.450.4035F} {450, 4035}

\bibitem[\protect\citeauthoryear{{Federrath}, {Banerjee}, {Clark}  \&
  {Klessen}}{{Federrath} et~al.}{2010}]{2010ApJ...713..269F}
{Federrath} C.,  {Banerjee} R.,  {Clark} P.~C.,   {Klessen} R.~S.,  2010,
  \mn@doi [\apj] {10.1088/0004-637X/713/1/269}, \href
  {https://ui.adsabs.harvard.edu/abs/2010ApJ...713..269F} {713, 269}

\bibitem[\protect\citeauthoryear{{Ferland}, {Peterson}, {Horne}, {Welsh}  \&
  {Nahar}}{{Ferland} et~al.}{1992}]{1992ApJ...387...95F}
{Ferland} G.~J.,  {Peterson} B.~M.,  {Horne} K.,  {Welsh} W.~F.,   {Nahar}
  S.~N.,  1992, \mn@doi [\apj] {10.1086/171063}, \href
  {https://ui.adsabs.harvard.edu/abs/1992ApJ...387...95F} {387, 95}

\bibitem[\protect\citeauthoryear{{Ferland}, {Korista}, {Verner}, {Ferguson},
  {Kingdon}  \& {Verner}}{{Ferland} et~al.}{1998}]{1998PASP..110..761F}
{Ferland} G.~J.,  {Korista} K.~T.,  {Verner} D.~A.,  {Ferguson} J.~W.,
  {Kingdon} J.~B.,   {Verner} E.~M.,  1998, \mn@doi [\pasp] {10.1086/316190},
  \href {https://ui.adsabs.harvard.edu/abs/1998PASP..110..761F} {110, 761}

\bibitem[\protect\citeauthoryear{{Fujita} et~al.,}{{Fujita}
  et~al.}{2021}]{2021PASJ...73S.201F}
{Fujita} S.,  et~al., 2021, \mn@doi [\pasj] {10.1093/pasj/psaa078}, \href
  {https://ui.adsabs.harvard.edu/abs/2021PASJ...73S.201F} {73, S201}

\bibitem[\protect\citeauthoryear{{Fukui} et~al.,}{{Fukui}
  et~al.}{1999}]{1999PASJ...51..745F}
{Fukui} Y.,  et~al., 1999, \mn@doi [\pasj] {10.1093/pasj/51.6.745}, \href
  {https://ui.adsabs.harvard.edu/abs/1999PASJ...51..745F} {51, 745}

\bibitem[\protect\citeauthoryear{{Fukui}, {Tsuge}, {Sano}, {Bekki}, {Yozin},
  {Tachihara}  \& {Inoue}}{{Fukui} et~al.}{2017}]{2017PASJ...69L...5F}
{Fukui} Y.,  {Tsuge} K.,  {Sano} H.,  {Bekki} K.,  {Yozin} C.,  {Tachihara} K.,
    {Inoue} T.,  2017, \mn@doi [\pasj] {10.1093/pasj/psx032}, \href
  {https://ui.adsabs.harvard.edu/abs/2017PASJ...69L...5F} {69, L5}

\bibitem[\protect\citeauthoryear{{Fukui} et~al.,}{{Fukui}
  et~al.}{2019}]{2019ApJ...886...14F}
{Fukui} Y.,  et~al., 2019, \mn@doi [\apj] {10.3847/1538-4357/ab4900}, \href
  {https://ui.adsabs.harvard.edu/abs/2019ApJ...886...14F} {886, 14}

\bibitem[\protect\citeauthoryear{{Fukui}, {Habe}, {Inoue}, {Enokiya}  \&
  {Tachihara}}{{Fukui} et~al.}{2021}]{2021PASJ...73S...1F}
{Fukui} Y.,  {Habe} A.,  {Inoue} T.,  {Enokiya} R.,   {Tachihara} K.,  2021,
  \mn@doi [\pasj] {10.1093/pasj/psaa103}, \href
  {https://ui.adsabs.harvard.edu/abs/2021PASJ...73S...1F} {73, S1}

\bibitem[\protect\citeauthoryear{{Geppert} et~al.,}{{Geppert}
  et~al.}{2005}]{2005JPhCS...4...26G}
{Geppert} W.~D.,  et~al., 2005, in Journal of Physics Conference Series. pp
  26--31, \mn@doi{10.1088/1742-6596/4/1/004}

\bibitem[\protect\citeauthoryear{{Glassgold} \& {Langer}}{{Glassgold} \&
  {Langer}}{1974}]{1974ApJ...193...73G}
{Glassgold} A.~E.,  {Langer} W.~D.,  1974, \mn@doi [\apj] {10.1086/153130},
  \href {https://ui.adsabs.harvard.edu/abs/1974ApJ...193...73G} {193, 73}

\bibitem[\protect\citeauthoryear{{Glover} \& {Clark}}{{Glover} \&
  {Clark}}{2012}]{2012MNRAS.421..116G}
{Glover} S. C.~O.,  {Clark} P.~C.,  2012, \mn@doi [\mnras]
  {10.1111/j.1365-2966.2011.20260.x}, \href
  {https://ui.adsabs.harvard.edu/abs/2012MNRAS.421..116G} {421, 116}

\bibitem[\protect\citeauthoryear{{Glover} \& {Jappsen}}{{Glover} \&
  {Jappsen}}{2007}]{2007ApJ...666....1G}
{Glover} S.~C.~O.,  {Jappsen} A.~K.,  2007, \mn@doi [\apj] {10.1086/519445},
  \href {https://ui.adsabs.harvard.edu/abs/2007ApJ...666....1G} {666, 1}

\bibitem[\protect\citeauthoryear{{Glover}, {Federrath}, {Mac Low}  \&
  {Klessen}}{{Glover} et~al.}{2010}]{2010MNRAS.404....2G}
{Glover} S.~C.~O.,  {Federrath} C.,  {Mac Low} M.~M.,   {Klessen} R.~S.,  2010,
  \mn@doi [\mnras] {10.1111/j.1365-2966.2009.15718.x}, \href
  {https://ui.adsabs.harvard.edu/abs/2010MNRAS.404....2G} {404, 2}

\bibitem[\protect\citeauthoryear{{Gong}, {Ostriker}  \& {Wolfire}}{{Gong}
  et~al.}{2017}]{2017ApJ...843...38G}
{Gong} M.,  {Ostriker} E.~C.,   {Wolfire} M.~G.,  2017, \mn@doi [\apj]
  {10.3847/1538-4357/aa7561}, \href
  {https://ui.adsabs.harvard.edu/abs/2017ApJ...843...38G} {843, 38}

\bibitem[\protect\citeauthoryear{{Grisdale}, {Agertz}, {Romeo}, {Renaud}  \&
  {Read}}{{Grisdale} et~al.}{2017}]{Grisdale2017}
{Grisdale} K.,  {Agertz} O.,  {Romeo} A.~B.,  {Renaud} F.,   {Read} J.~I.,
  2017, \mn@doi [\mnras] {10.1093/mnras/stw3133}, \href
  {https://ui.adsabs.harvard.edu/abs/2017MNRAS.466.1093G} {466, 1093}

\bibitem[\protect\citeauthoryear{{Habe} \& {Ohta}}{{Habe} \&
  {Ohta}}{1992}]{1992PASJ...44..203H}
{Habe} A.,  {Ohta} K.,  1992, \pasj, \href
  {https://ui.adsabs.harvard.edu/abs/1992PASJ...44..203H} {44, 203}

\bibitem[\protect\citeauthoryear{{Habing}}{{Habing}}{1968}]{1968BAN....19..421H}
{Habing} H.~J.,  1968, \bain, \href
  {https://ui.adsabs.harvard.edu/abs/1968BAN....19..421H} {19, 421}

\bibitem[\protect\citeauthoryear{{Haid}, {Walch}, {Seifried}, {W{\"u}nsch},
  {Dinnbier}  \& {Naab}}{{Haid} et~al.}{2018}]{Haid18}
{Haid} S.,  {Walch} S.,  {Seifried} D.,  {W{\"u}nsch} R.,  {Dinnbier} F.,
  {Naab} T.,  2018, \mn@doi [\mnras] {10.1093/mnras/sty1315}, \href
  {https://ui.adsabs.harvard.edu/abs/2018MNRAS.478.4799H} {478, 4799}

\bibitem[\protect\citeauthoryear{Harris et~al.,}{Harris
  et~al.}{2020}]{harris2020array}
Harris C.~R.,  et~al., 2020, \mn@doi [Nature] {10.1038/s41586-020-2649-2}, 585,
  357

\bibitem[\protect\citeauthoryear{{Haworth}, {Shima}, {Tasker}, {Fukui},
  {Torii}, {Dale}, {Takahira}  \& {Habe}}{{Haworth} et~al.}{2015}]{Haworth2015}
{Haworth} T.~J.,  {Shima} K.,  {Tasker} E.~J.,  {Fukui} Y.,  {Torii} K.,
  {Dale} J.~E.,  {Takahira} K.,   {Habe} A.,  2015, \mn@doi [\mnras]
  {10.1093/mnras/stv2068}, \href
  {https://ui.adsabs.harvard.edu/abs/2015MNRAS.454.1634H} {454, 1634}

\bibitem[\protect\citeauthoryear{{Heays}, {Bosman}  \& {van Dishoeck}}{{Heays}
  et~al.}{2017}]{2017A&A...602A.105H}
{Heays} A.~N.,  {Bosman} A.~D.,   {van Dishoeck} E.~F.,  2017, \mn@doi [\aap]
  {10.1051/0004-6361/201628742}, \href
  {https://ui.adsabs.harvard.edu/abs/2017A&A...602A.105H} {602, A105}

\bibitem[\protect\citeauthoryear{{Hollenbach} \& {McKee}}{{Hollenbach} \&
  {McKee}}{1979}]{1979ApJS...41..555H}
{Hollenbach} D.,  {McKee} C.~F.,  1979, \mn@doi [\apjs] {10.1086/190631}, \href
  {https://ui.adsabs.harvard.edu/abs/1979ApJS...41..555H} {41, 555}

\bibitem[\protect\citeauthoryear{{Hu}, {Federrath}, {Xu}  \& {Mathew}}{{Hu}
  et~al.}{2022}]{Hu2022}
{Hu} Y.,  {Federrath} C.,  {Xu} S.,   {Mathew} S.~S.,  2022, arXiv e-prints,
  \href {https://ui.adsabs.harvard.edu/abs/2022arXiv220301508H} {p.
  arXiv:2203.01508}

\bibitem[\protect\citeauthoryear{{Hummer} \& {Storey}}{{Hummer} \&
  {Storey}}{1998}]{1998MNRAS.297.1073H}
{Hummer} D.~G.,  {Storey} P.~J.,  1998, \mn@doi [\mnras]
  {10.1046/j.1365-8711.1998.2970041073.x}, \href
  {https://ui.adsabs.harvard.edu/abs/1998MNRAS.297.1073H} {297, 1073}

\bibitem[\protect\citeauthoryear{Hunter}{Hunter}{2007}]{Hunter:2007}
Hunter J.~D.,  2007, \mn@doi [Computing in Science \& Engineering]
  {10.1109/MCSE.2007.55}, 9, 90

\bibitem[\protect\citeauthoryear{{Jacobs}, {Giedt}  \& {Cohen}}{{Jacobs}
  et~al.}{1967}]{1967JChPh..47...54J}
{Jacobs} T.~A.,  {Giedt} R.~R.,   {Cohen} N.,  1967, \mn@doi [\jcp]
  {10.1063/1.1711890}, \href
  {https://ui.adsabs.harvard.edu/abs/1967JChPh..47...54J} {47, 54}

\bibitem[\protect\citeauthoryear{{Janev}, {Langer}  \& {Evans}}{{Janev}
  et~al.}{1987}]{1987ephh.book.....J}
{Janev} R.~K.,  {Langer} W.~D.,   {Evans} K.,  1987, {Elementary processes in
  Hydrogen-Helium plasmas - Cross sections and reaction rate coefficients}.
Springer

\bibitem[\protect\citeauthoryear{{Jones}, {Clark}, {Glover}  \&
  {Hacar}}{{Jones} et~al.}{2021}]{Jones_etal_2022}
{Jones} G.~H.,  {Clark} P.~C.,  {Glover} S. C.~O.,   {Hacar} A.,  2021, MNRAS,
  submitted, \href {https://ui.adsabs.harvard.edu/abs/2021arXiv211205543J} {p.
  arXiv:2112.05543}

\bibitem[\protect\citeauthoryear{{Karpas}, {Anicich}  \& {Huntress}}{{Karpas}
  et~al.}{1979}]{1979JChPh..70.2877K}
{Karpas} Z.,  {Anicich} V.,   {Huntress} W.~T.,  1979, \mn@doi [\jcp]
  {10.1063/1.437823}, \href
  {https://ui.adsabs.harvard.edu/abs/1979JChPh..70.2877K} {70, 2877}

\bibitem[\protect\citeauthoryear{{Kim}, {Theard}  \& {Huntress}}{{Kim}
  et~al.}{1975}]{1975CPL....32..610K}
{Kim} J.~K.,  {Theard} L.~P.,   {Huntress} W.~T. J.,  1975, \mn@doi [Chemical
  Physics Letters] {10.1016/0009-2614(75)85252-3}, \href
  {https://ui.adsabs.harvard.edu/abs/1975CPL....32..610K} {32, 610}

\bibitem[\protect\citeauthoryear{{Kippenhahn}, {Weigert}  \&
  {Weiss}}{{Kippenhahn} et~al.}{2012}]{Kippenhahn12}
{Kippenhahn} R.,  {Weigert} A.,   {Weiss} A.,  2012, {Stellar Structure and
  Evolution}, \mn@doi{10.1007/978-3-642-30304-3.
}

\bibitem[\protect\citeauthoryear{{Krumholz} \& {McKee}}{{Krumholz} \&
  {McKee}}{2005}]{2005ApJ...630..250K}
{Krumholz} M.~R.,  {McKee} C.~F.,  2005, \mn@doi [\apj] {10.1086/431734}, \href
  {https://ui.adsabs.harvard.edu/abs/2005ApJ...630..250K} {630, 250}

\bibitem[\protect\citeauthoryear{{Lada}, {Lombardi}  \& {Alves}}{{Lada}
  et~al.}{2010}]{Lada2010}
{Lada} C.~J.,  {Lombardi} M.,   {Alves} J.~F.,  2010, \mn@doi [\apj]
  {10.1088/0004-637X/724/1/687}, \href
  {https://ui.adsabs.harvard.edu/abs/2010ApJ...724..687L} {724, 687}

\bibitem[\protect\citeauthoryear{{Lepp} \& {Shull}}{{Lepp} \&
  {Shull}}{1983}]{1983ApJ...270..578L}
{Lepp} S.,  {Shull} J.~M.,  1983, \mn@doi [\apj] {10.1086/161149}, \href
  {https://ui.adsabs.harvard.edu/abs/1983ApJ...270..578L} {270, 578}

\bibitem[\protect\citeauthoryear{{Leroy} et~al.}{{Leroy}
  et~al.}{2021}]{Leroy2021}
{Leroy} A.~K.,  et~al., 2021, arXiv e-prints, \href
  {https://ui.adsabs.harvard.edu/abs/2021arXiv210407739L} {p. arXiv:2104.07739}

\bibitem[\protect\citeauthoryear{{Linder}, {Janev}  \& {Botero}}{{Linder}
  et~al.}{1995}]{1995ampf.conf..397L}
{Linder} F.,  {Janev} R.~K.,   {Botero} J.,  1995, in {Janev} R.~K.,  ed.,
  Atomic and Molecular Processes in Fusion Edge Plasmas. p.~397

\bibitem[\protect\citeauthoryear{{Liow} \& {Dobbs}}{{Liow} \&
  {Dobbs}}{2020}]{Liow2020}
{Liow} K.~Y.,  {Dobbs} C.~L.,  2020, \mn@doi [\mnras] {10.1093/mnras/staa2857},
  \href {https://ui.adsabs.harvard.edu/abs/2020MNRAS.499.1099L} {499, 1099}

\bibitem[\protect\citeauthoryear{{Mac Low} \& {Shull}}{{Mac Low} \&
  {Shull}}{1986}]{1986ApJ...302..585M}
{Mac Low} M.~M.,  {Shull} J.~M.,  1986, \mn@doi [\apj] {10.1086/164017}, \href
  {https://ui.adsabs.harvard.edu/abs/1986ApJ...302..585M} {302, 585}

\bibitem[\protect\citeauthoryear{{Martin}, {Schwarz}  \& {Mandy}}{{Martin}
  et~al.}{1996}]{1996ApJ...461..265M}
{Martin} P.~G.,  {Schwarz} D.~H.,   {Mandy} M.~E.,  1996, \mn@doi [\apj]
  {10.1086/177053}, \href
  {https://ui.adsabs.harvard.edu/abs/1996ApJ...461..265M} {461, 265}

\bibitem[\protect\citeauthoryear{{Martin}, {Keogh}  \& {Mandy}}{{Martin}
  et~al.}{1998}]{1998ApJ...499..793M}
{Martin} P.~G.,  {Keogh} W.~J.,   {Mandy} M.~E.,  1998, \mn@doi [\apj]
  {10.1086/305665}, \href
  {https://ui.adsabs.harvard.edu/abs/1998ApJ...499..793M} {499, 793}

\bibitem[\protect\citeauthoryear{{Matthaeus}, {Pouquet}, {Mininni}, {Dmitruk}
  \& {Breech}}{{Matthaeus} et~al.}{2008}]{Matthaeus2008}
{Matthaeus} W.~H.,  {Pouquet} A.,  {Mininni} P.~D.,  {Dmitruk} P.,   {Breech}
  B.,  2008, \mn@doi [\prl] {10.1103/PhysRevLett.100.085003}, \href
  {https://ui.adsabs.harvard.edu/abs/2008PhRvL.100h5003M} {100, 085003}

\bibitem[\protect\citeauthoryear{{McCall} et~al.,}{{McCall}
  et~al.}{2004}]{2004PhRvA..70e2716M}
{McCall} B.~J.,  et~al., 2004, \mn@doi [\pra] {10.1103/PhysRevA.70.052716},
  \href {https://ui.adsabs.harvard.edu/abs/2004PhRvA..70e2716M} {70, 052716}

\bibitem[\protect\citeauthoryear{{McElroy}, {Walsh}, {Markwick}, {Cordiner},
  {Smith}  \& {Millar}}{{McElroy} et~al.}{2013}]{2013A&A...550A..36M}
{McElroy} D.,  {Walsh} C.,  {Markwick} A.~J.,  {Cordiner} M.~A.,  {Smith} K.,
  {Millar} T.~J.,  2013, \mn@doi [\aap] {10.1051/0004-6361/201220465}, \href
  {https://ui.adsabs.harvard.edu/abs/2013A&A...550A..36M} {550, A36}

\bibitem[\protect\citeauthoryear{{McKinnon}, {Torrey}, {Vogelsberger},
  {Hayward}  \& {Marinacci}}{{McKinnon} et~al.}{2017}]{2017MNRAS.468.1505M}
{McKinnon} R.,  {Torrey} P.,  {Vogelsberger} M.,  {Hayward} C.~C.,
  {Marinacci} F.,  2017, \mn@doi [\mnras] {10.1093/mnras/stx467}, \href
  {https://ui.adsabs.harvard.edu/abs/2017MNRAS.468.1505M} {468, 1505}

\bibitem[\protect\citeauthoryear{{Mitchell}}{{Mitchell}}{1984}]{1984ApJS...54...81M}
{Mitchell} G.~F.,  1984, \mn@doi [\apjs] {10.1086/190919}, \href
  {https://ui.adsabs.harvard.edu/abs/1984ApJS...54...81M} {54, 81}

\bibitem[\protect\citeauthoryear{{Motte}, {Bontemps}  \& {Louvet}}{{Motte}
  et~al.}{2018}]{2018ARA&A..56...41M}
{Motte} F.,  {Bontemps} S.,   {Louvet} F.,  2018, \mn@doi [\araa]
  {10.1146/annurev-astro-091916-055235}, \href
  {https://ui.adsabs.harvard.edu/abs/2018ARA&A..56...41M} {56, 41}

\bibitem[\protect\citeauthoryear{{Nahar}}{{Nahar}}{2000}]{2000ApJS..126..537N}
{Nahar} S.~N.,  2000, \mn@doi [\apjs] {10.1086/313307}, \href
  {https://ui.adsabs.harvard.edu/abs/2000ApJS..126..537N} {126, 537}

\bibitem[\protect\citeauthoryear{{Neelamkodan}, {Tokuda}, {Barman}, {Kondo},
  {Sano}  \& {Onishi}}{{Neelamkodan} et~al.}{2021}]{2021ApJ...908L..43N}
{Neelamkodan} N.,  {Tokuda} K.,  {Barman} S.,  {Kondo} H.,  {Sano} H.,
  {Onishi} T.,  2021, \mn@doi [\apjl] {10.3847/2041-8213/abdebb}, \href
  {https://ui.adsabs.harvard.edu/abs/2021ApJ...908L..43N} {908, L43}

\bibitem[\protect\citeauthoryear{{Nelson} \& {Langer}}{{Nelson} \&
  {Langer}}{1999}]{1999ApJ...524..923N}
{Nelson} R.~P.,  {Langer} W.~D.,  1999, \mn@doi [\apj] {10.1086/307823}, \href
  {https://ui.adsabs.harvard.edu/abs/1999ApJ...524..923N} {524, 923}

\bibitem[\protect\citeauthoryear{{Osterbrock}}{{Osterbrock}}{1989}]{1989agna.book.....O}
{Osterbrock} D.~E.,  1989, {Astrophysics of gaseous nebulae and active galactic
  nuclei}.
University Science Books

\bibitem[\protect\citeauthoryear{Pakmor \& Springel}{Pakmor \&
  Springel}{2013}]{10.1093/mnras/stt428}
Pakmor R.,  Springel V.,  2013, \mn@doi [\mnras] {10.1093/mnras/stt428}, 432,
  176

\bibitem[\protect\citeauthoryear{{Pequignot} \& {Aldrovandi}}{{Pequignot} \&
  {Aldrovandi}}{1986}]{1986A&A...161..169P}
{Pequignot} D.,  {Aldrovandi} S.~M.~V.,  1986, \aap, \href
  {https://ui.adsabs.harvard.edu/abs/1986A&A...161..169P} {161, 169}

\bibitem[\protect\citeauthoryear{{Petuchowski}, {Dwek}, {Allen}  \&
  {Nuth}}{{Petuchowski} et~al.}{1989}]{1989ApJ...342..406P}
{Petuchowski} S.~J.,  {Dwek} E.,  {Allen} J.~E. J.,   {Nuth} J.~A. I.,  1989,
  \mn@doi [\apj] {10.1086/167601}, \href
  {https://ui.adsabs.harvard.edu/abs/1989ApJ...342..406P} {342, 406}

\bibitem[\protect\citeauthoryear{{Powell}, {Roe}, {Linde}, {Gombosi}  \& {De
  Zeeuw}}{{Powell} et~al.}{1999}]{1999JCoPh.154..284P}
{Powell} K.~G.,  {Roe} P.~L.,  {Linde} T.~J.,  {Gombosi} T.~I.,   {De Zeeuw}
  D.~L.,  1999, \mn@doi [J.\ Comp.\ Phys.] {10.1006/jcph.1999.6299}, \href
  {https://ui.adsabs.harvard.edu/abs/1999JCoPh.154..284P} {154, 284}

\bibitem[\protect\citeauthoryear{{Prasad} \& {Huntress}}{{Prasad} \&
  {Huntress}}{1980}]{1980ApJS...43....1P}
{Prasad} S.~S.,  {Huntress} W.~T. J.,  1980, \mn@doi [\apjs] {10.1086/190665},
  \href {https://ui.adsabs.harvard.edu/abs/1980ApJS...43....1P} {43, 1}

\bibitem[\protect\citeauthoryear{{Prole}, {Clark}, {Klessen}  \&
  {Glover}}{{Prole} et~al.}{2022}]{Prole_etal_2022}
{Prole} L.~R.,  {Clark} P.~C.,  {Klessen} R.~S.,   {Glover} S. C.~O.,  2022,
  \mn@doi [\mnras] {10.1093/mnras/stab3697}, \href
  {https://ui.adsabs.harvard.edu/abs/2022MNRAS.510.4019P} {510, 4019}

\bibitem[\protect\citeauthoryear{{Rahner}, {Pellegrini}, {Glover}  \&
  {Klessen}}{{Rahner} et~al.}{2017}]{Rahner17}
{Rahner} D.,  {Pellegrini} E.~W.,  {Glover} S. C.~O.,   {Klessen} R.~S.,  2017,
  \mn@doi [\mnras] {10.1093/mnras/stx1532}, \href
  {https://ui.adsabs.harvard.edu/abs/2017MNRAS.470.4453R} {470, 4453}

\bibitem[\protect\citeauthoryear{{Rahner}, {Pellegrini}, {Glover}  \&
  {Klessen}}{{Rahner} et~al.}{2019}]{Rahner19}
{Rahner} D.,  {Pellegrini} E.~W.,  {Glover} S. C.~O.,   {Klessen} R.~S.,  2019,
  \mn@doi [\mnras] {10.1093/mnras/sty3295}, \href
  {https://ui.adsabs.harvard.edu/abs/2019MNRAS.483.2547R} {483, 2547}

\bibitem[\protect\citeauthoryear{{Roman-Duval}, {Jackson}, {Heyer}, {Rathborne}
   \& {Simon}}{{Roman-Duval} et~al.}{2010}]{Duval2014}
{Roman-Duval} J.,  {Jackson} J.~M.,  {Heyer} M.,  {Rathborne} J.,   {Simon} R.,
   2010, \mn@doi [\apj] {10.1088/0004-637X/723/1/492}, \href
  {https://ui.adsabs.harvard.edu/abs/2010ApJ...723..492R} {723, 492}

\bibitem[\protect\citeauthoryear{{Scoville}, {Sanders}  \&
  {Clemens}}{{Scoville} et~al.}{1986}]{1986ApJ...310L..77S}
{Scoville} N.~Z.,  {Sanders} D.~B.,   {Clemens} D.~P.,  1986, \mn@doi [\apjl]
  {10.1086/184785}, \href
  {https://ui.adsabs.harvard.edu/abs/1986ApJ...310L..77S} {310, L77}

\bibitem[\protect\citeauthoryear{{Sembach}, {Howk}, {Ryans}  \&
  {Keenan}}{{Sembach} et~al.}{2000}]{2000ApJ...528..310S}
{Sembach} K.~R.,  {Howk} J.~C.,  {Ryans} R. S.~I.,   {Keenan} F.~P.,  2000,
  \mn@doi [\apj] {10.1086/308173}, \href
  {https://ui.adsabs.harvard.edu/abs/2000ApJ...528..310S} {528, 310}

\bibitem[\protect\citeauthoryear{{Shapiro} \& {Kang}}{{Shapiro} \&
  {Kang}}{1987}]{1987ApJ...318...32S}
{Shapiro} P.~R.,  {Kang} H.,  1987, \mn@doi [\apj] {10.1086/165350}, \href
  {https://ui.adsabs.harvard.edu/abs/1987ApJ...318...32S} {318, 32}

\bibitem[\protect\citeauthoryear{{Shore}}{{Shore}}{1981}]{Shore1981}
{Shore} S.~N.,  1981, \mn@doi [\apj] {10.1086/159263}, \href
  {https://ui.adsabs.harvard.edu/abs/1981ApJ...249...93S} {249, 93}

\bibitem[\protect\citeauthoryear{{Skarbinski}, {Jeffreson}  \&
  {Goodman}}{{Skarbinski} et~al.}{2022}]{Skarbinski2022}
{Skarbinski} M.~S.,  {Jeffreson} S. M.~R.,   {Goodman} A.~A.,  2022, arXiv
  e-prints, \href {https://ui.adsabs.harvard.edu/abs/2022arXiv221201396S} {p.
  arXiv:2212.01396}

\bibitem[\protect\citeauthoryear{{Smith} et~al.,}{{Smith}
  et~al.}{2017}]{2017MNRAS.466.2217S}
{Smith} B.~D.,  et~al., 2017, \mn@doi [\mnras] {10.1093/mnras/stw3291}, \href
  {https://ui.adsabs.harvard.edu/abs/2017MNRAS.466.2217S} {466, 2217}

\bibitem[\protect\citeauthoryear{{Soler} \& {Hennebelle}}{{Soler} \&
  {Hennebelle}}{2017}]{Soler2017}
{Soler} J.~D.,  {Hennebelle} P.,  2017, \mn@doi [\aap]
  {10.1051/0004-6361/201731049}, \href
  {https://ui.adsabs.harvard.edu/abs/2017A&A...607A...2S} {607, A2}

\bibitem[\protect\citeauthoryear{{Sormani} et~al.,}{{Sormani}
  et~al.}{2019}]{2019MNRAS.488.4663S}
{Sormani} M.~C.,  et~al., 2019, \mn@doi [\mnras] {10.1093/mnras/stz2054}, \href
  {https://ui.adsabs.harvard.edu/abs/2019MNRAS.488.4663S} {488, 4663}

\bibitem[\protect\citeauthoryear{{Springel}}{{Springel}}{2005}]{2005MNRAS.364.1105S}
{Springel} V.,  2005, \mn@doi [\mnras] {10.1111/j.1365-2966.2005.09655.x},
  \href {https://ui.adsabs.harvard.edu/abs/2005MNRAS.364.1105S} {364, 1105}

\bibitem[\protect\citeauthoryear{{Springel}}{{Springel}}{2010}]{2010MNRAS.401..791S}
{Springel} V.,  2010, \mn@doi [\mnras] {10.1111/j.1365-2966.2009.15715.x},
  \href {https://ui.adsabs.harvard.edu/abs/2010MNRAS.401..791S} {401, 791}

\bibitem[\protect\citeauthoryear{{Stancil}, {Lepp}  \& {Dalgarno}}{{Stancil}
  et~al.}{1998}]{1998ApJ...509....1S}
{Stancil} P.~C.,  {Lepp} S.,   {Dalgarno} A.,  1998, \mn@doi [\apj]
  {10.1086/306473}, \href
  {https://ui.adsabs.harvard.edu/abs/1998ApJ...509....1S} {509, 1}

\bibitem[\protect\citeauthoryear{{Stancil}, {Schultz}, {Kimura}, {Gu}, {Hirsch}
   \& {Buenker}}{{Stancil} et~al.}{1999}]{1999A&AS..140..225S}
{Stancil} P.~C.,  {Schultz} D.~R.,  {Kimura} M.,  {Gu} J.~P.,  {Hirsch} G.,
  {Buenker} R.~J.,  1999, \mn@doi [\aaps] {10.1051/aas:1999419}, \href
  {https://ui.adsabs.harvard.edu/abs/1999A&AS..140..225S} {140, 225}

\bibitem[\protect\citeauthoryear{{Takahira}, {Tasker}  \& {Habe}}{{Takahira}
  et~al.}{2014}]{Takahira2014}
{Takahira} K.,  {Tasker} E.~J.,   {Habe} A.,  2014, \mn@doi [\apj]
  {10.1088/0004-637X/792/1/63}, \href
  {https://ui.adsabs.harvard.edu/abs/2014ApJ...792...63T} {792, 63}

\bibitem[\protect\citeauthoryear{{Tan}, {Beltr{\'a}n}, {Caselli}, {Fontani},
  {Fuente}, {Krumholz}, {McKee}  \& {Stolte}}{{Tan}
  et~al.}{2014}]{2014prpl.conf..149T}
{Tan} J.~C.,  {Beltr{\'a}n} M.~T.,  {Caselli} P.,  {Fontani} F.,  {Fuente} A.,
  {Krumholz} M.~R.,  {McKee} C.~F.,   {Stolte} A.,  2014, in {Beuther} H.,
  {Klessen} R.~S.,  {Dullemond} C.~P.,   {Henning} T.,  eds, Protostars and
  Planets VI. p.~149 (\mn@eprint {arXiv} {1402.0919}),
  \mn@doi{10.2458/azu\_uapress\_9780816531240-ch007}

\bibitem[\protect\citeauthoryear{{Tanvir} \& {Dale}}{{Tanvir} \&
  {Dale}}{2020}]{2020MNRAS.494..246T}
{Tanvir} T.~S.,  {Dale} J.~E.,  2020, \mn@doi [\mnras] {10.1093/mnras/staa665},
  \href {https://ui.adsabs.harvard.edu/abs/2020MNRAS.494..246T} {494, 246}

\bibitem[\protect\citeauthoryear{{Tasker} \& {Tan}}{{Tasker} \&
  {Tan}}{2009}]{2009ApJ...700..358T}
{Tasker} E.~J.,  {Tan} J.~C.,  2009, \mn@doi [\apj]
  {10.1088/0004-637X/700/1/358}, \href
  {https://ui.adsabs.harvard.edu/abs/2009ApJ...700..358T} {700, 358}

\bibitem[\protect\citeauthoryear{{Tokuda} et~al.,}{{Tokuda}
  et~al.}{2019}]{2019ApJ...886...15T}
{Tokuda} K.,  et~al., 2019, \mn@doi [\apj] {10.3847/1538-4357/ab48ff}, \href
  {https://ui.adsabs.harvard.edu/abs/2019ApJ...886...15T} {886, 15}

\bibitem[\protect\citeauthoryear{{Trevisan} \& {Tennyson}}{{Trevisan} \&
  {Tennyson}}{2002}]{2002PPCF...44.1263T}
{Trevisan} C.~S.,  {Tennyson} J.,  2002, \mn@doi [Plasma Physics and Controlled
  Fusion] {10.1088/0741-3335/44/7/315}, \href
  {https://ui.adsabs.harvard.edu/abs/2002PPCF...44.1263T} {44, 1263}

\bibitem[\protect\citeauthoryear{{Tsang} \& {Hampson}}{{Tsang} \&
  {Hampson}}{1986}]{1986JPCRD..15.1087T}
{Tsang} W.,  {Hampson} R.~F.,  1986, \mn@doi [Journal of Physical and Chemical
  Reference Data] {10.1063/1.555759}, \href
  {https://ui.adsabs.harvard.edu/abs/1986JPCRD..15.1087T} {15, 1087}

\bibitem[\protect\citeauthoryear{{Vissapragada}, {Buzard}, {Miller},
  {O'Connor}, {de Ruette}, {Urbain}  \& {Savin}}{{Vissapragada}
  et~al.}{2016}]{2016ApJ...832...31V}
{Vissapragada} S.,  {Buzard} C.~F.,  {Miller} K.~A.,  {O'Connor} A.~P.,  {de
  Ruette} N.,  {Urbain} X.,   {Savin} D.~W.,  2016, \mn@doi [\apj]
  {10.3847/0004-637X/832/1/31}, \href
  {https://ui.adsabs.harvard.edu/abs/2016ApJ...832...31V} {832, 31}

\bibitem[\protect\citeauthoryear{{Voronov}}{{Voronov}}{1997}]{1997ADNDT..65....1V}
{Voronov} G.~S.,  1997, \mn@doi [Atomic Data and Nuclear Data Tables]
  {10.1006/adnd.1997.0732}, \href
  {https://ui.adsabs.harvard.edu/abs/1997ADNDT..65....1V} {65, 1}

\bibitem[\protect\citeauthoryear{{Wakelam} et~al.,}{{Wakelam}
  et~al.}{2010}]{2010SSRv..156...13W}
{Wakelam} V.,  et~al., 2010, \mn@doi [\ssr] {10.1007/s11214-010-9712-5}, \href
  {https://ui.adsabs.harvard.edu/abs/2010SSRv..156...13W} {156, 13}

\bibitem[\protect\citeauthoryear{{Wareing}, {Pittard}  \& {Falle}}{{Wareing}
  et~al.}{2017}]{2017MNRAS.470.2283W}
{Wareing} C.~J.,  {Pittard} J.~M.,   {Falle} S.~A.~E.~G.,  2017, \mn@doi
  [\mnras] {10.1093/mnras/stx1417}, \href
  {https://ui.adsabs.harvard.edu/abs/2017MNRAS.470.2283W} {470, 2283}

\bibitem[\protect\citeauthoryear{{Weingartner} \& {Draine}}{{Weingartner} \&
  {Draine}}{2001}]{wd01}
{Weingartner} J.~C.,  {Draine} B.~T.,  2001, \mn@doi [\apj] {10.1086/324035},
  \href {https://ui.adsabs.harvard.edu/abs/2001ApJ...563..842W} {563, 842}

\bibitem[\protect\citeauthoryear{{Wolfire}, {Tielens}, {Hollenbach}  \&
  {Kaufman}}{{Wolfire} et~al.}{2008}]{2008ApJ...680..384W}
{Wolfire} M.~G.,  {Tielens} A.~G.~G.~M.,  {Hollenbach} D.,   {Kaufman} M.~J.,
  2008, \mn@doi [\apj] {10.1086/587688}, \href
  {https://ui.adsabs.harvard.edu/abs/2008ApJ...680..384W} {680, 384}

\bibitem[\protect\citeauthoryear{{Wollenberg}, {Glover}, {Clark}  \&
  {Klessen}}{{Wollenberg} et~al.}{2020}]{Wollenberg2020}
{Wollenberg} K. M.~J.,  {Glover} S. C.~O.,  {Clark} P.~C.,   {Klessen} R.~S.,
  2020, \mn@doi [\mnras] {10.1093/mnras/staa289}, \href
  {https://ui.adsabs.harvard.edu/abs/2020MNRAS.494.1871W} {494, 1871}

\bibitem[\protect\citeauthoryear{{Wu}, {Van Loo}, {Tan}  \& {Bruderer}}{{Wu}
  et~al.}{2015}]{2015ApJ...811...56W}
{Wu} B.,  {Van Loo} S.,  {Tan} J.~C.,   {Bruderer} S.,  2015, \mn@doi [\apj]
  {10.1088/0004-637X/811/1/56}, \href
  {https://ui.adsabs.harvard.edu/abs/2015ApJ...811...56W} {811, 56}

\bibitem[\protect\citeauthoryear{{Wu}, {Tan}, {Christie}, {Nakamura}, {Van Loo}
   \& {Collins}}{{Wu} et~al.}{2017}]{2017ApJ...841...88W}
{Wu} B.,  {Tan} J.~C.,  {Christie} D.,  {Nakamura} F.,  {Van Loo} S.,
  {Collins} D.,  2017, \mn@doi [\apj] {10.3847/1538-4357/aa6ffa}, \href
  {https://ui.adsabs.harvard.edu/abs/2017ApJ...841...88W} {841, 88}

\bibitem[\protect\citeauthoryear{{Zanchet}, {Bussery-Honvault}, {Jorfi}  \&
  {Honvault}}{{Zanchet} et~al.}{2009}]{2009PCCP...11.6182Z}
{Zanchet} A.,  {Bussery-Honvault} B.,  {Jorfi} M.,   {Honvault} P.,  2009,
  \mn@doi [Physical Chemistry Chemical Physics (Incorporating Faraday
  Transactions)] {10.1039/b903829a}, \href
  {https://ui.adsabs.harvard.edu/abs/2009PCCP...11.6182Z} {11, 6182}

\bibitem[\protect\citeauthoryear{{de Ruette}, {Miller}, {O'Connor}, {Urbain},
  {Buzard}, {Vissapragada}  \& {Savin}}{{de Ruette}
  et~al.}{2016}]{2016ApJ...816...31D}
{de Ruette} N.,  {Miller} K.~A.,  {O'Connor} A.~P.,  {Urbain} X.,  {Buzard}
  C.~F.,  {Vissapragada} S.,   {Savin} D.~W.,  2016, \mn@doi [\apj]
  {10.3847/0004-637X/816/1/31}, \href
  {https://ui.adsabs.harvard.edu/abs/2016ApJ...816...31D} {816, 31}

\bibitem[\protect\citeauthoryear{{van Dishoeck}}{{van
  Dishoeck}}{1988}]{1988ASSL..146...49V}
{van Dishoeck} E.~F.,  1988, in {Millar} T.~J.,  {Williams} D.~A.,  eds, ,
  Vol.~146, Rate Coefficients in Astrochemistry.
Kluwer, p.~49, \mn@doi{10.1007/978-94-009-3007-0\_4}

\makeatother
\end{thebibliography}

%\begin{thebibliography}{99}
%\end{thebibliography}

%%%%%%%%%%%%%%%%%%%%%%%%%%%%%%%%%%%%%%%%%%%%%%%%%%

%%%%%%%%%%%%%%%%% APPENDICES %%%%%%%%%%%%%%%%%%%%%

\appendix

\section{Chemical network}
\label{app:chem}
The reactions included in our chemical network are summarized in Tables~\ref{tab:coll} and \ref{tab:other}. Our network is based on the one presented by \citeauthor{2017ApJ...843...38G}~(2017; hereafter, G17), but also includes several reactions not included in their network, denoted in the tables as ``Not in G17''. In addition, for some reactions we have adopted a different rate coefficient from the one listed in G17; these are also indicated in the Table, with the note ``Different rate''. Reactions with no attached note are treated exactly the same as in G17. Below, we discuss the rationale for the differences between our network and the original G17 network. For the most part, these differences are to make the network more robust when applied to physical conditions outside of the range considered by G17 and have little impact on its behaviour in the typical photodissociation region (PDR) conditions that were the main focus of their study.

\begin{table}
\caption{List of collisional chemical reactions included in our chemical network \label{tab:coll}}
\begin{tabular}{cllc}
No.\ & Reaction & Notes & Refs.\  \\
\hline
1 & ${\rm H + e^{-}}  \rightarrow {\rm H^+ + e^- + e^-}$ & & 1 \\
2 & ${\rm H^+ + e} \rightarrow {\rm H + \gamma}$ & & 2 \\
3 &   ${\rm He^+ + H_2} \rightarrow {\rm H_2^+ + He}$ & & 3 \\
4 &  ${\rm He^+ + H_2} \rightarrow {\rm H^+ + H + He}$ & & 3 \\
5 &  ${\rm H_2 + e^{-}} \rightarrow {\rm H + H + e^{-}}$ & Not in G17 & 4 \\
6 &  ${\rm H_2 + H} \rightarrow {\rm H + H + H}$ & Different rate & 5, 6, 7 \\
7 &  ${\rm H_2 + H_2} \rightarrow {\rm H + H + H_2}$ & Different rate & 8, 9 \\
8 &  ${\rm H_2^+ + H_2} \rightarrow {\rm H_3^+ + H}$ & Different rate & 10 \\
9 &  ${\rm H_2^+ + H}  \rightarrow {\rm H_2 + H^+}$ & & 11 \\
10 &  ${\rm H_3^+ + e^-} \rightarrow {\rm H + H + H}$ & & 12 \\
11 & ${\rm H_3^+ + e^-} \rightarrow {\rm H_2 + H}$ & & 12 \\
12 & ${\rm He + e^{-}}  \rightarrow {\rm He^+ + e^- + e^-}$ & Not in G17 & 1 \\
13 & ${\rm He^+ + e^{-}} \rightarrow {\rm He + \gamma}$ & Different rate & 13, 14 \\ %-- Gong et al just use case B rate, we try to account for coupling with H
14 & ${\rm C^+ + H_2} \rightarrow {\rm CH_x + H}$ & & 15, 16 \\
15 & ${\rm C^+ + H_2 + e^{-}} \rightarrow {\rm C + H + H}$ & & 15, 16 \\
16 & ${\rm C + H_2}  \rightarrow {\rm CH_x + \gamma}$ & Not in G17 & 17 \\
17 & ${\rm C + H_3^+} \rightarrow {\rm CH_x + H_2}$ & & 18 \\  % --- with rate from erratum, not original article
18 & ${\rm C^+ + e^{-}} \rightarrow {\rm C + \gamma}$ & Different rate & 14, 19 \\
19 & ${\rm C + e^{-}} \rightarrow {\rm C^+ + e^{-} + e^{-}}$ & Not in G17 & 20 \\
20 & ${\rm O^+ + H}  \rightarrow {\rm O + H^+}$ & & 21 \\
21 & ${\rm O  + H^+} \rightarrow {\rm O^+ + H}$ & & 21  \\
22 & ${\rm O^+ + H_2} \rightarrow {\rm OH_x + H}$ & & 16 \\
23 & ${\rm O^+ + H_2 + e} \rightarrow {\rm O + H + H}$ & & 16  \\
24 & ${\rm O  + H_3^+} \rightarrow {\rm OH_x + H_2}$ & & 22  \\
25 & ${\rm O  + H_3^+ + e^{-}} \rightarrow {\rm H_2 + O + H}$ & & 22  \\
26 & ${\rm C^+ + OH_x} \rightarrow {\rm HCO^+}$ & & 16 \\
27 & ${\rm C  + OH_x} \rightarrow {\rm CO + H}$ & & 23 \\
28 & ${\rm CH_x + He^+} \rightarrow {\rm C^+ + He + H}$ & Not in G17 & 17, 24 \\
29 & ${\rm CH_x + H} \rightarrow {\rm H_2 + C}$ & & 15 \\
30 & ${\rm CH_x + O} \rightarrow {\rm CO + H}$ & & 15 \\
31 & ${\rm OH_x + H} \rightarrow {\rm O + H_2}$ & Not in G17 & 24, 25 \\
32 & ${\rm OH_x + O} \rightarrow {\rm O + O + H}$ & & 26 \\
33 & ${\rm OH_x + He^+} \rightarrow {\rm O^{+} + He + H}$ & & 15 \\
34 & ${\rm  CO + H_3^+} \rightarrow {\rm HCO^+ + H_2}$ & & 27 \\
35 & ${\rm CO + He^+} \rightarrow {\rm C^+ + O + He}$ & Different rate & 28 \\
36 & ${\rm CO + H} \rightarrow {\rm C + OH_x}$ & Not in G17 & 29 \\
37 & ${\rm HCO^+ + e^-} \rightarrow {\rm OH_x + C}$ & Not in G17 & 30 \\
38 & ${\rm HCO^+ + e^-} \rightarrow {\rm CO + H}$ & & 30 \\
39 & ${\rm Si^+ + e^-} \rightarrow {\rm Si + \gamma}$ & Different rate & 31 \\
40 & ${\rm Si  + e^-} \rightarrow {\rm Si^+ + e^- + e^-}$ & Not in G17 & 20 \\
\hline
\end{tabular}
\medskip
\\ {\bf References:} 
1: \citet{1987ephh.book.....J}; 
2: \citet{1992ApJ...387...95F}; 
3: \citet{1984PhDT.......142B};
4: \citet{2002PPCF...44.1263T};
5: \citet{1986ApJ...302..585M};
6: \citet{1983ApJ...270..578L};
7: \citet{1996ApJ...461..265M};
8: \citet{1998ApJ...499..793M};
9: \citet{1967JChPh..47...54J};
10: \citet{1995ampf.conf..397L};
11: \citet{1979JChPh..70.2877K};
12: \citet{2004PhRvA..70e2716M};
13: \citet{1998MNRAS.297.1073H};
14: \citet{2006ApJS..167..334B};
15: \citet{2010SSRv..156...13W};
16: \citet{2017ApJ...843...38G};
17: \citet{1980ApJS...43....1P};
18: \citet{2016ApJ...832...31V};
19: \citet{2003A&A...406.1151B};
20: \citet{1997ADNDT..65....1V};
21: \citet{1999A&AS..140..225S};
22: Fit by G17 to \citet{2016ApJ...816...31D};
23: \citet{2009PCCP...11.6182Z};
24: \citet{2013A&A...550A..36M}
25: \citet{1986JPCRD..15.1087T};
26: \citet{2006JPCA..110.3101C};
27: \citet{1975CPL....32..610K};
28: \citet{1989ApJ...342..406P};
29: \citet{1984ApJS...54...81M};
30: \citet{2005JPhCS...4...26G};
31: \citet{2000ApJS..126..537N};
\end{table}

\begin{table}
\caption{List of grain surface, cosmic ray and photochemical reactions included in our chemical network \label{tab:other}}
\begin{tabular}{cllc}
No.\ & Reaction & Notes & Refs.\  \\
\hline
41 & ${\rm H + H + gr} \rightarrow {\rm H_{2} + gr}$  & Different rate & 1 \\
42 & ${\rm H^{+} + e^{-} + gr} \rightarrow {\rm H + gr}$  & Different rate  & 2 \\
43 & ${\rm C^{+} + e^- + gr} \rightarrow {\rm C + gr}$  & Different rate  & 2 \\
44 & ${\rm He^{+} + e^- + gr} \rightarrow {\rm He + gr}$  & Different rate  & 2 \\
45 & ${\rm Si^{+} + e^- + gr} \rightarrow {\rm Si + gr}$  & Different rate  & 2 \\
\hline
46 & ${\rm H + cr} \rightarrow {\rm H^{+} + e^{-}}$ & & 3 \\
47 & ${\rm H_{2} + cr} \rightarrow {\rm H_{2}^{+} + e^{-}}$ & & 3 \\
48 & ${\rm H_{2} + cr} \rightarrow {\rm H + H^{+} + e^{-}}$ & Not in G17 & 4 \\
49 & ${\rm H_{2} + cr} \rightarrow {\rm H + H}$ & Not in G17 & 4 \\
50 & ${\rm He + cr} \rightarrow {\rm He^{+} + e^{-}}$ & & 4 \\
51 & ${\rm C + cr} \rightarrow {\rm C^{+} + e^{-}}$ & & 4 \\
52 & ${\rm CO + cr + H} \rightarrow {\rm HCO^{+} + e^{-}}$ & & 5 \\
53 & ${\rm C + \gamma_{\rm cr}} \rightarrow {\rm C^{+} + e^{-}}$ & & 6 \\
54 & ${\rm CO + \gamma_{\rm cr}} \rightarrow {\rm C^{+} + e^{-}}$ & & 6 \\
55 & ${\rm Si + \gamma_{\rm cr}} \rightarrow {\rm Si^{+} + e^{-}}$ & & 6 \\
\hline
56 & ${\rm C + \gamma \rightarrow C^{+} + e^{-}}$ & & 6 \\
57 & ${\rm CH_x + \gamma \rightarrow C + H}$ & & 5,6 \\
58 & ${\rm CO + \gamma \rightarrow C + O}$ & & 6 \\
59 & ${\rm OH_x + \gamma \rightarrow O + H}$ & & 5,6 \\
60 & ${\rm Si + \gamma \rightarrow Si^{+} + e^{-}}$ & & 7 \\
61 & ${\rm H_2 + \gamma \rightarrow H + H}$ & & 8 \\
\hline
\end{tabular}
\medskip
\\ 
The primary cosmic ray ionization rate of atomic hydrogen is a free parameter in our chemical model and the value we select for this in our simulations is discussed in the main text. The total (primary plus secondary) rate for H (reaction 46), as well as the total cosmic ray ionization rates of H$_{2}$, He, C, CO and Si (reactions 47--55) are scaled relative to this value using scaling factors derived from the cited references. \\
 {\bf References:} 1: \citet{1979ApJS...41..555H}; 2: \citet{wd01}, modified as described in the text; 3: \citet{1974ApJ...193...73G}; 4: \citet{2013A&A...550A..36M}; 5: \citet{2017ApJ...843...38G}; 6: \citet{2017A&A...602A.105H}; 7: \citet{1988ASSL..146...49V}; 8: 1996ApJ...468..269D
\end{table}

\subsection*{Reactions in our network that are not in G17}

\subsubsection*{Reactions 5, 12, 19 \& 40}
These reactions -- the collisional dissociation of H$_{2}$ by electrons and the collisional ionisation of He, C and Si -- were neglected by G17 because they are unimportant at typical PDR temperatures. However, they can become important in hot shocked gas with $T \gg 10^{4}$~K and we include them to ensure that the chemical network behaves reasonably at these high temperatures. 

\subsubsection*{Reaction 16}
The formation of CH$_{\rm x}$ -- a pseudo-molecule that represents  light hydrocarbons such as CH, CH$_{2}$, CH$^{+}$ etc.\ -- by radiative association of atomic carbon and H$_{2}$ is neglected by G17 because it is a slow process and in typical Milky Way conditions is unimportant compared to CH$_{x}$ formation via the reaction of C and H$_{3}^{+}$ (reaction 17). However, the abundance of H$_3^{+}$, and hence the rate of reaction 17, depends sensitively on the cosmic ray ionisation rate, and so reaction 16 can become important in conditions where this is much smaller than the typical Milky Way value.

\subsubsection*{Reaction 28}
In neutral gas, this reaction is unimportant compared to reaction 29. It is included here to ensure that an appropriate loss route exists for CH$_{\rm x}$ in very highly ionized gas. For the rate coefficient, we use the value given by \citet{2013A&A...550A..36M} for the reaction
\begin{equation*}
{\rm CH + He^{+}} \rightarrow {\rm C^{+} + He + H},
\end{equation*}
which comes originally from \citet{1980ApJS...43....1P}.

\subsubsection*{Reaction 31}
This reaction is unimportant compared to reaction 32 at temperatures lower than 500~K, but quickly becomes dominant at higher temperatures, given a sufficient supply of hydrogen atoms. It is therefore not important in typical PDR conditions but can become important in shocks. For the rate coefficient, we adopt the value given by \citet{2013A&A...550A..36M} for the reaction
\begin{equation*}
{\rm OH + H} \rightarrow {\rm O + H_{2}},
\end{equation*}
which is based on \citet{1986JPCRD..15.1087T}.

\subsubsection*{Reaction 36}
This reaction has a substantial activation energy ($E/k \sim 78000~{\rm K}$) and is therefore unimportant in typical PDR conditions. However, it can become important in hot, shocked gas. In particular, we have found that if strong shocks occur in gas with high $A_{\rm V}$, CO can persist in the gas up to artificially high temperatures of $> 10^{4}$~K if this reaction is not included.

\subsubsection*{Reaction 37}
As G17 note in their appendix A, the destruction of HCO$^{+}$ is dominated by the reaction ${\rm HCO^{+} + e^{-} \rightarrow CO + H}$ (reaction 38 above). However, although CO and H are the most likely products of the dissociative recombination of HCO$^{+}$, roughly 8\% of the time this process instead yields C and OH \citep{2005JPhCS...4...26G}. We include this outcome here for completeness.

\subsubsection*{Reactions 48 \& 49}
Although interactions between high energy cosmic rays and H$_{2}$ molecules primarily produce H$_{2}^{+}$ ions (reaction 47), a small fraction of the time the outcome can instead be a hydrogen atom, a proton and an electron (reaction 48) or two hydrogen atoms (reaction 49). These outcomes were neglected by G17 but we include them here for completeness. 

\subsection*{Reactions with different rate coefficients}

\subsubsection*{Reactions 6 \& 7}
We use the same low density limits for the rates of these reactions as in G17, and the same expression for the H$_{2}$ critical density. However, we use slightly different expressions for the high density limits: G17 follow \citet{1983ApJ...270..578L} and \citet{1987ApJ...318...32S} for reactions 6 and 7, respectively, whereas we use the expression given by \citet{1996ApJ...461..265M} for the high density limit of reaction 6 and adopt a value 8 times smaller than this for reaction 7, following \citet{1967JChPh..47...54J}. At densities below the H$_{2}$ critical density ($n \sim 10^{4} \: {\rm cm^{-3}}$), we therefore recover the same behaviour as in G17, and we also find good agreement between the different treatments in hot ($T > 6000$~K), high density gas. The only significant difference comes in cool dense gas, where the \citet{1983ApJ...270..578L} expression over-estimates the H$_{2}$ collisional dissociation rate. However, given the small size of this rate at these temperatures, this difference is likely only of minor importance.

\subsubsection*{Reaction 8}
We adopt the rate coefficient for this reaction given in \citet{1998ApJ...509....1S}, which is their fit to cross-section data from \citet{1995ampf.conf..397L}. G17 also cite \citet{1995ampf.conf..397L} as the source of their rate coefficient, but their expression is $\sim$27\% larger than the one given in \citet{1998ApJ...509....1S}. The source of this discrepancy is unclear. However, in practice it is unlikely to be important as this reaction is never the rate-limiting step for the formation of H$_{3}^{+}$.

\subsubsection*{Reaction 13}
G17 assume case B for the radiative recombination of He$^{+}$ and use a reaction rate coefficient from \citet{2010MNRAS.404....2G} that is a fit to the values tabulated by \citet{1998MNRAS.297.1073H}. Our treatment differs from this in two respects. First, in addition to radiative recombination, we also account for dielectronic recombination of He$^{+}$, using a rate coefficient from \citet{2006ApJS..167..334B}. Second, although we assume the on-the-spot approximation applies, we do not assume pure case B recombination for He$^{+}$, which would be valid only in a gas consisting of pure helium. Instead, we follow \citet{1989agna.book.....O} and account for the fact that some of the photons produced during the recombination of He$^{+}$ are absorbed by atomic hydrogen rather than He. (A more detailed discussion of how this is done can be found in \citealt{2007ApJ...666....1G}).

\subsubsection*{Reaction 18}
As in G17, we adopt a rate coefficient for C$^{+}$ recombination that is the sum of two contributions: one corresponding to radiative recombination, taken from \citet{2006ApJS..167..334B}, and one corresponding to dielectronic recombination, taken from \citet{2003A&A...406.1151B}. In the expression that they use for the dielectronic recombination rate, G17 retain only the first three terms, which is sufficient at low temperatures but which leads to inaccuracies at high temperatures ($T \gg 10^{4}$~K). In our implementation of this rate, we instead retain all of the terms from the expression given by \citet{2003A&A...406.1151B}.

\subsubsection*{Reaction 35}
G17 adopt a temperature-independent rate for this reaction from \citet{1986ApJS...62..553A}, whereas we adopt the temperature-dependent value proposed by \citet{1989ApJ...342..406P}. In practice, there is very little difference between these two values at typical PDR temperatures.

\subsubsection*{Reaction 39}
G17 adopt a rate coefficient for this reaction that they credit to \citet{2013A&A...550A..36M} but that derives originally from \citet{1986A&A...161..169P}. However, this fit is formally valid only in the temperature range $10 < T < 1000$~K. Moreover, it only accounts for the contribution from radiative recombination, and not the dielectronic recombination term that dominates at high temperatures. We adopt instead a rate from \citet{2000ApJS..126..537N} that accounts for both processes.

\subsubsection*{Reaction 41}
G17 adopt a constant value of $3 \times 10^{-17} \: {\rm cm^{3} \: s^{-1}}$  for the rate coefficient for this reaction. We instead adopt the rate coefficient given in \citet{1979ApJS...41..555H}, which depends on the temperatures of both the gas and the dust grains.  

\subsubsection*{Reaction 42--45}
We follow G17 in that we use the reaction rate coefficients given in \citet{wd01} for the recombination of H$^{+}$, C$^{+}$, He$^{+}$ and Si$^{+}$ ions on grain surfaces (reactions 42--45), multiplied by a factor of 0.6 to better match the results of \citet{2008ApJ...680..384W}. These rate coefficients depend primarily on the parameter
\begin{equation}
\psi = \frac{G \sqrt{T}}{n_{\rm e}},
\end{equation}
where $G$ is the local value of the interstellar radiation field in \citet{1968BAN....19..421H}
units and $n_{\rm e}$ is the electron number density. One important way in which our treatment 
differs from G17 is in our treatment of these rate coefficients for low values of $\psi$. The expressions given in \citet{wd01} are stated to be valid only for $\psi > 100 \: {\rm K^{1/2} \: cm^{3}}$ and applying them unaltered when the value of $\psi$ is smaller than this yields recombination rates that are significant over-estimates of the true values. To avoid this, we simply assume that the rates in gas with $\psi < 100 \: {\rm K^{1/2} \: cm^{3}}$ are the same as those in gas with $\psi = 100 \: {\rm K^{1/2} \: cm^{3}}$ (c.f.\ figure 3 in \citealt{wd01}).

The other main difference between our treatment and that in G17 is that we multiply the grain surface recombination rates by an additional factor of $\exp(-T / 34000)$. This is to ensure that the recombination rates fall rapidly to zero in very hot gas, in conditions where we expect that in reality the dust would be quickly destroyed by sputtering. This modification would not be necessary if we were using a more sophisticated treatment of dust evolution that accounted for this effect \citep[see e.g.][]{2017MNRAS.468.1505M}, but this is a topic for future work.

\section{Post-collision magnetic field}
\label{sect:mag}
\begin{figure*}
    \centering
    \includegraphics[width=0.9\textwidth]{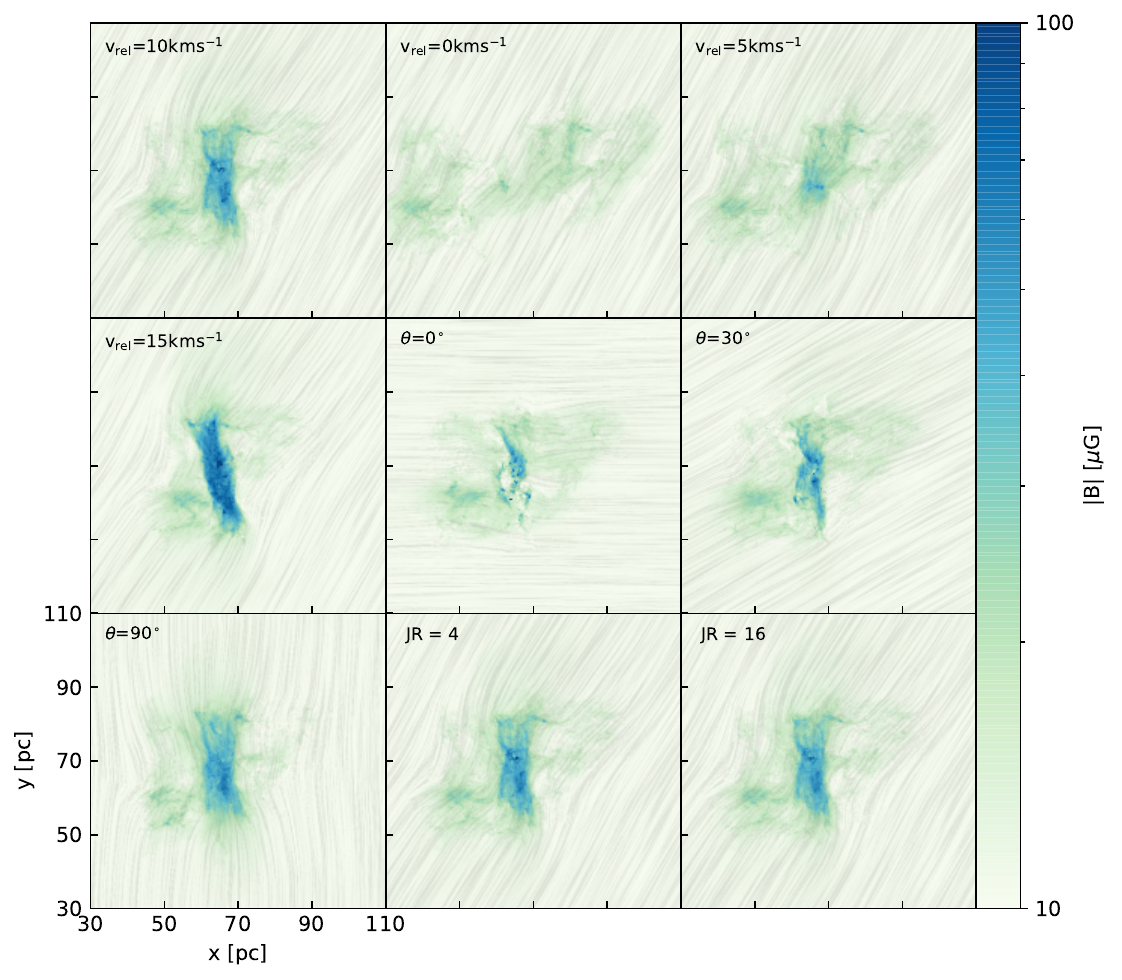}
    \caption{Maps of the magnetic field strength of the simulations carried out at $t=2.40$ Myr. Line integral convolution (LIC) is used on the map to indicate the direction of the magnetic field in the $x$-$y$ plane.}
    \label{fig:mag_abs}
\end{figure*}

\begin{figure*}
    \centering
    \includegraphics[width=0.9\textwidth]{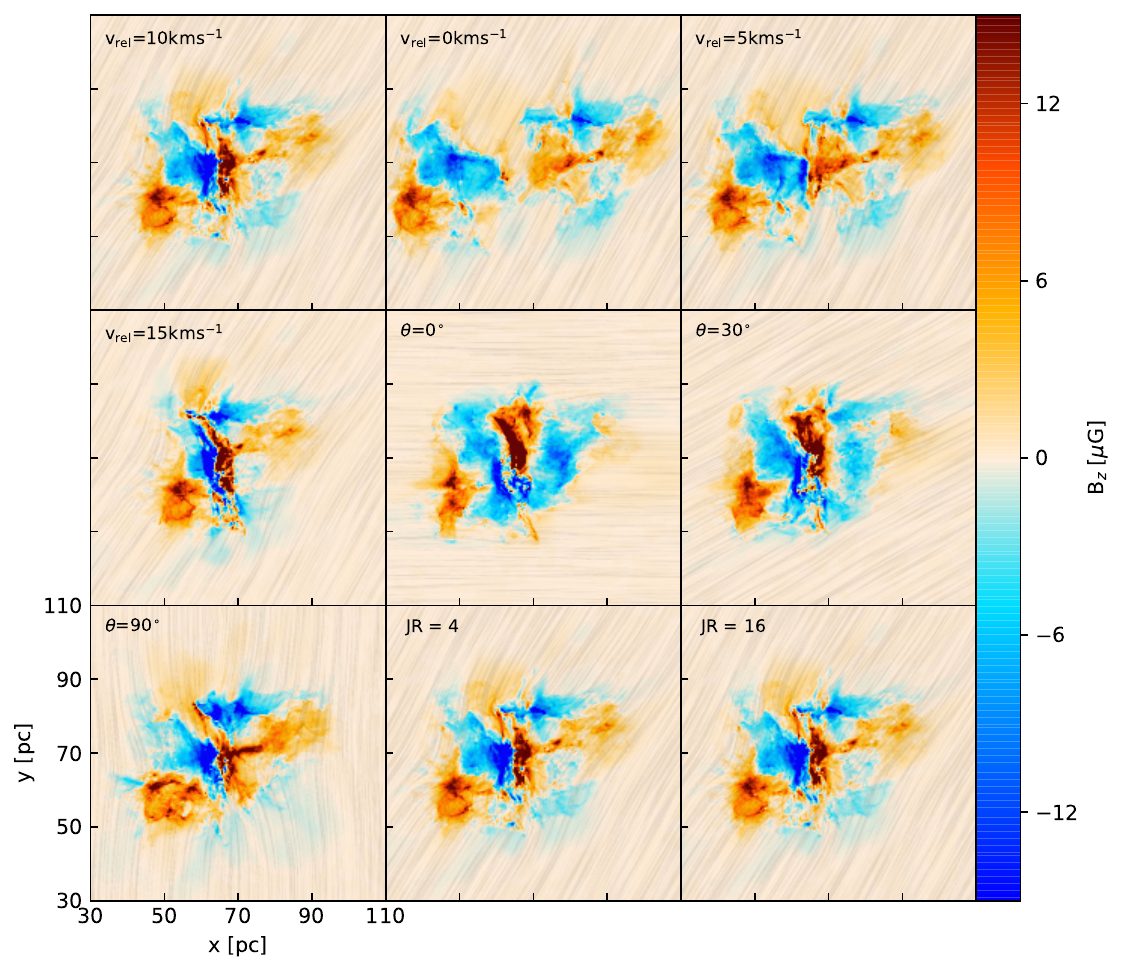}
    \caption{Same as figure~\ref{fig:mag_abs} but for the z component of the magnetic field}
    \label{fig:mag_z}
\end{figure*}

Here we illustrate how the magnetic field is warped in the process of the collision of the clouds. Figures~\ref{fig:mag_abs} \& \ref{fig:mag_z} show images of the magnitude and $z$-component of the magnetic field, respectively, which have been convolved with the $x$-$y$ components of magnetic field via line integral convolution (LIC) \citep{10.1145/166117.166151}. This is done to give indication of the direction of the magnetic field in the $x$-$y$ plane.

We noted in Section~\ref{sect:SFR} that an early onset of star formation is observed for a magnetic field inclination of $\theta = 0$, i.e.\ a field that is initially parallel to the collision axis. In this case, the magnetic field is not compressed during the collision and hence does not hinder the collision process. This can be seen clearly in the
central panels of Figures~\ref{fig:mag_abs} \& \ref{fig:mag_z}, where the pattern of the map generated by the LIC lies predominantly parallel to the $x$-axis. Only at the site of the collision is the magnetic field distorted, owing to the local collapse of the gas.

%%%%%%%%%%%%%%%%%%%%%%%%%%%%%%%%%%%%%%%%%%%%%%%%%%

% Don't change these lines
\bsp	% typesetting comment
\label{lastpage}
\end{document}